\documentclass[onecollarge]{svjour2}       
\usepackage{graphicx}
\usepackage{alltt}

\begin{document}

\title{Tidal synchronization of close-in satellites and exoplanets. A rheophysical approach. }
\author{Sylvio Ferraz-Mello}
\institute{Instituto de Astronomia 
Geof\'{\i}sica e Ci\^encias Atmosf\'ericas\\Universidade de S\~ao Paulo, Brasil \\
sylvio [at] usp.br}
\titlerunning{Tidal Synchronization}

\maketitle

\begin{abstract}
This paper presents a new theory of the dynamical tides of celestial bodies. 
It is founded on a Newtonian creep instead of the classical delaying approach of the standard viscoelastic theories and the results of the theory derive mainly from the solution of a non-homogeneous ordinary differential equation. 
Lags appear in the solution but as quantities determined from the solution of the equation and are not arbitrary external quantities plugged in an elastic model. 
The resulting lags of the tide components are increasing functions of their frequencies (as in Darwin's theory), but not small quantities. 
The amplitudes of the tide components depend on the viscosity of the body and on their frequencies; they are not constants. 
The resulting stationary rotations (often called pseudo-synchronous) have an excess velocity roughly proportional to $6ne^2/(\chi^2+\chi^{-2})$ ($\chi$ is the mean-motion in units of one critical frequency - the relaxation factor - inversely proportional to the viscosity) instead of the exact $6ne^2$ of standard theories. The dissipation in the pseudo-synchronous solution is inversely proportional to $(\chi+\chi^{-1})$; thus, in the inviscid limit, it is roughly proportional to the frequency (as in standard theories), but that behavior is inverted when the viscosity is high and the tide frequency larger than the critical frequency. 
For free rotating bodies, the dissipation is given by the same law, but now $\chi$ is the frequency of the semi-diurnal tide in units of the critical frequency. This approach fails, however, to reproduce the actual tidal lags on Earth. In this case, to reconcile theory and observations, we need to assume the existence of an elastic tide superposed to the creeping tide. 
The theory is applied to several Solar System and extrasolar bodies and currently available data are used to estimate the relaxation factor $\gamma$ (i.e. the critical frequency) of these bodies.
\end{abstract}

\def\beq{\begin{equation}}
\def\endeq{\end{equation}}
\def\begdi{\begin{displaymath}}
\def\enddi{\end{displaymath}}
\def\ep{\varepsilon}
\def\epp{\varepsilon^\prime}
\def\eppp{\varepsilon^{\prime\prime}}
\def\App{A^\prime}
\def\Appp{A^{\prime\prime}}
\def\bpp{\beta^\prime}
\def\bppp{\beta^{\prime\prime}}
\def\Cpp{C^\prime}
\def\Cppp{C^{\prime\prime}}
\def\CppQ{C^{\prime 2}}
\def\CpppQ{C^{\prime\prime 2}}
\def\aprmaior{\;\buildrel\hbox{\geq}\over{\sim}\;}    
\def\defeq{\;\buildrel\hbox{\small def}\over{\,=}\;}    
\def\speq{\hspace{1mm} = \hspace{1mm}}    
\def\FRH{Ferraz-Mello et al (2008)}
\def\hilight{\textbf}

\section{Introduction}

During the $20^{th}$ century, many versions of the Darwin theory, or of what has been called 
Darwin theory, were used in the study of the tidal evolution of satellites and planets (see reviews in Ogilvie and Lin, 2004, Efroimsky and Williams, 2009). Those versions were not exempt of problems, the more important appearing when they are used to study solutions near the spin-orbit resonance. All of them, consistently, show the existence of a stationary solution, which is synchronous when the two bodies move in circular orbits or super-synchronous otherwise (that is, a solution in which the time average of the rotation angular velocity is constant and slightly higher than the orbital mean motion; it is often called pseudo-synchronous). 
The excess of angular velocity in the stationary solution when the bodies are in elliptical orbits is physically expected. The torques acting on the bodies are inversely proportional to a great power of the distance and, therefore, much larger when the body is at the pericenter of the relative orbit than in other parts of the orbit. As a consequence, the angular velocity near the pericenter will enter in the time averages with a larger weight and will dominate the result leading to averages larger than the mean motion $n$. Then, in the case of a planet or satellite moving around its primary in an elliptic orbit, we should not expect the synchronization of the two motions. 
Pure tidal theories lead to synchronous stationary solutions only in the circular approximation.
In standard Darwin theories\footnote{I will have to refer often in this paper to theories derived from Darwin's theory. 
To use a simple label, I will use the word ``standard Darwin theories", or, for short, ``standard theories", 
to denote all Darwin-like theories in which an elastic tide is delayed by lags assumed small and proportional to frequencies.}, in the usual approximation in which only the main zonal harmonic is considered, the average excess of angular velocity of the body is given by $\sim 6ne^2$  ($e$ is the orbit eccentricity)(See Goldreich and Peale, 1966, eqn. 24). This is a quantity independent of the nature of the body and is thus one important difficulty of these theories. 

This prediction is, however, not confirmed by the observation of planetary satellites. 
Titan, for example, should then have a synodic rotation period of about 8.5 years (i.e. $\sim 43^\circ$ per year) while the radar observations done with the space probe Cassini over several years showed that the actual rotation differs from the synchronous spin by a shift of $\sim 0.12^\circ$ per year in apparent longitude (Stiles et al. 2008, 2010)\footnote{A re-analysis of the data by Meriggiola and Iess (Meriggiola, 2012) has not showed discrepancy from a synchronous motion larger than $0.02^\circ$ per year}. 
In the case of Europa, the value predicted by the standard theory is less than 20 years. The present position of some cycloidal cracks confirms a non-synchronous rotation, but the comparison of Voyager and Galileo images indicate a synodic period less than  12,000 years (Hoppa et al. 1999, Greenberg et al. 2002, Hurford et al. 2007). 
The most striking case is the Moon whose average rotation and orbit are synchronous notwithstanding an orbital eccentricity 0.055.

The only way to conciliate theory and observation is to assume that in all these cases an extra torque able to counterbalance the tidal torque is acting on the body, and the most obvious assumption concerning this extra torque is the existence of a permanent equatorial asymmetry of the body (Greenberg and Weidenschilling, 1984). This can explain the case of the Moon and a quick calculation including the tides raised by the Earth and the $C_{31}$ component of the lunar potential result in a spin-orbit synchronous solution; the net effect of the tides is just a deviation of the symmetry axis which is not pointing to the Earth but shows a small offset (see \FRH, eqn 46). In the case of Titan and Europa, the situation is more complex. In both cases, we may assume a permanent equatorial asymmetry, but the resulting solution is then an exact synchronization, not a slightly non-synchronous rotation as some observations seem to show.

Several attempts were made to explain these differences either by assuming ad hoc mass distributions inside these bodies, or by modifying the Darwinian theories.
Since the $6ne^2$ law results from the theory when the lags are assumed to be small and proportional to the frequency of the tide components, one immediate idea is to substitute the linear dependence by a power law (see Sears et al. 1993). However, without physical grounds to support the assumption, the result will depend on the \textit{ad hoc} fixed powers and remain only speculative. Efroimsky and Lainey (2007) have proposed to substitute the linear law by an inverse power law. The grounds for their proposal are some laboratory measurements and also the determination of the dissipation affecting the Earth's seismic waves at different frequencies. 
An inverse power law brings with it an additional difficulty because any quantity inversely proportional to the frequency tends to infinity when the frequency goes to zero.
Efroimsky and Williams (2009) and Efroimsky (2012) claim that this difficulty can be circumvented, but it is done at the price of an extremely complex modeling. The results of this paper (Sec. \ref{sec:lag}.1)
may help the understanding of what happens in the immediate neighborhood of the frequency zero and why no actual singularity exists.

A different approach was presented in \FRH (hereafter FRH) in which the lags remain proportional to the frequencies, but the non-instantaneous response of the body to the tidal potential is taken into account. In that approach, the resulting excess of angular velocity is given by $\sim 6ne^2 (k_1/k_0)$ where $k_0$ and $k_1$ denote the response factors of the body to the semi-diurnal and the monthly components of the tide, respectively\footnote{We use in this paper the same tide component names used for fast rotating bodies (Type I of FRH) regardless of its actual rotation speed. In next sections  we use the names monthly and/or annual for tidal components with the same period as the orbital motion.}. These response factors are not equal. In the stationary condition, the frequency of the semi-diurnal component approaches 0 and $k_0$ approaches its maximum, the fluid Love number $k_f$. 
On its turn, the response of the monthly component of the tide will depend on the viscosity of the body. If the viscosity is small, the body will respond faster and $k_1/k_0 \sim 1$. The result is again $\sim 6ne^2$. However, if the viscosity is large, the deformation of the body does not attain its maximum theoretical extent; that is,  $k_1/k_0 < 1$ and the resulting excess of angular velocity  is smaller than $6ne^2$. However, as in the discussion above, without physical grounds to support the chosen value of $k_1$, the result will depend on \textit{ad hoc} fixed values. 

A difference in the response factors was also considered by Darwin (1880), but it was only sporadically been considered in some papers (e.g. Alexander, 1973; Wahr, 1981; Efroimsky and Williams, 2009). However, the differences among their response factors were not sufficient to solve the problem highlighted above, mainly because those differences disappear when the so-called ``weak friction approximation" (Alexander, 1973) is introduced. 

One structural difficulty comes from the use of Love's theory of elasticity. The use of Love's theorem as a shortcut to obtain the potential of the field spanned by the tidally deformed body without having to calculate beforehand its figure of equilibrium introduces the constant Love numbers as response factors for all terms issued from the same spherical harmonic of the tidal potential.
The only free parameters are, then, the \textit{ad hoc} introduced phase lags. 
After Darwin (1879), the tangents of the phase lags are proportional to the frequencies of the tide components, one result also found in this paper. 
However, since the lags are introduced in standard theories as arbitrary \textit{ad hoc} quantities, different laws can be postulated. 
They can be freely postulated or, as in some more elaborated investigations, suggested from the study of delays in damped oscillators (see, for instance, Greenberg, 2010). 

However, in most of the standard theories, the viscosity of the body is never explicitly considered and strictly speaking the so-called ``viscoelastic" approaches are more adequately described as ``elastic" and ``delayed", their actual ingredients. 

We did not consider in this short account theories based on energy dissipation instead of phase delays, because they follow a different line of thought (for a comprehensive account of them, see Eggleton, 2006, chap. 4, and Migaszewkski, 2012. See also Bambusi and Haus, 2012). The results issued from these theories are formally equivalent to Darwin's theory when lags are assumed proportional to the frequencies and lead to the same pseudo-synchronous results discussed in the beginning of this Introduction.

This paper introduces a new rheophysical approach in which the body tends always to creep towards the equilibrium by the only action of the gravitational forces acting on it (self-gravitation and tidal potential) and does it with a rate inversely proportional to its viscosity. The adopted creep law is Newtonian (linear), and at every instant the stress is assumed to be proportional to the distance from the equilibrium. This leads to eq. (\ref{eq:ansatz}) and with only one exception every result in the paper is a direct consequence of this first-order differential equation. 
The exception occurs when studying the shape of the tide in hard bodies as the Earth. In this case, in order to conciliate theory and observation, it is necessary to assume that a purely elastic tide (See Sec. \ref{sec:lag}) exists, superposed to the creep tide. 

The physical model is presented in Section 2 and developed in sections 3 and 4.
Section  2 also presents a short application to the case of bodies in circular motion with results equal to those obtained by Darwin (1879) and which served as the basis for the introduction of a lag proportional to the tide frequency in his 1880 theory. The next sections are devoted to the calculation of the perturbations: First, the perturbations on the rotation of the tidally deformed body (Sec. 5) and its synchronization (or pseudo-synchronization) (Sec. 6); then, the perturbations on the semi-major axis (energy dissipation; Sec. 7) and on the eccentricity (Sec.8). Section 9 discusses the value of the relaxation factor $\gamma$ on the basis of our current knowledge of the tidal evolution of stars, planets and planetary satellites. The creeping tide theory is completed, in Section 10, by the introduction of an additional elastic tide, necessary to reproduce the observed shape of tidally deformed bodies.

\section{A simple rheophysical model}
The usual standard model considers an elastic tide and delays the tidal bulge by an \textit {ad hoc} phase lag in order to take into account the body anelasticity. 
In this theory, we propose instead of that, a simple rheophysical model, the basis of which is shown in fig. 1.  
We consider one body of mass $m$ and assume that, at a given time $t$, the surface of the body is a function $\zeta=\zeta(\widehat\varphi^*,\widehat\theta^*,t)$ where $\zeta$ is the distance of the surface points to the center of gravity of the body and $\widehat\varphi^*$, $\widehat\theta^*$ their longitudes and co-latitudes with respect to a reference system rotating with the body. In the same instant $t$, the body is under the action of a tidal potential due to one second body of mass $M$ situated in its neighborhood. No hypothesis is being done on the relative importance of the two bodies. Both may play the role of the central body and its satellite or hot planet. In actual applications, both cases have to be considered (see FRH Sec. 18).

Would the body $\tens{m}$ be inviscid, it would immediately change its shape to the equilibrium configuration. In the simplest case, the figure of equilibrium of it under the action of the tidal potential is a prolate Jeans spheroid $\rho=\rho(\widehat\varphi^*,\widehat\theta^*,t)$ (see Chandrasekhar 1969) whose major axis is directed along the line joining the centers of gravity of the two bodies. If $a_e,b_e$ are the principal equatorial axes of the spheroid, its prolateness is
\begin{equation}\label{prola}
\epsilon_\rho=\frac{a_e}{b_e}-1=\frac{15}{4}\Bigg{(}\frac{M}{m}\Bigg{)}\Bigg{(}\frac{R_e}{r}\Bigg{)}^3
\end{equation}
(Tisserand, 1891) where $R_e$ is the mean equatorial radius of $\tens{m}$ and $r$ the distance from $\tens{M}$ to $\tens{m}$. Terms of second order with respect to $\epsilon_\rho$ are neglected in this and in the following calculations.

\begin{figure}[t]
\centerline{\hbox{
\includegraphics[height=5cm,clip=]{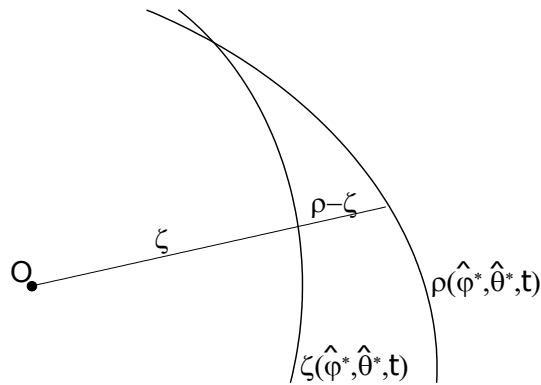}}}
\caption{Elements of the model: $\zeta$ is a section of the surface of the body at the time $t$; $\rho$ is a section of the surface of the equilibrium spheroid at the same time.}
\label{ansatz}       
\end{figure}

The adopted model is founded on the law
\beq
\dot{\zeta} = \gamma (\rho-\zeta).
\label{eq:ansatz}\endeq
The basic idea supporting this law is that because of the forces acting on the body (self-gravitation plus tide), its surface will tend to the equilibrium spheroid, but not instantaneously, and its instantaneous response (measured by $\dot{\zeta}$) will be proportional to the radial separation between it and the equilibrium spheroid, $\rho-\zeta$. Eq. (\ref{eq:ansatz}) is the equation of a \textit{Newtonian creep} (see Oswald, 2009, chap. 5) where the distance to the equilibrium was considered as proportional to the stress. 
It does not consider inertia or azimuthal motions, which may exist and should be considered in further studies.
The superposed elastic tide whose existence stems from the comparison of the observed shape of the tidal deformations and the theory, is not included in the equations of the model because the force due to the elastic tide is radial and its torque is zero; it does not affect the rotation or the averaged work and eccentricity. 
 
The relaxation factor $\gamma$ is a radial deformation rate gradient and has dimension {\sl{\rm T}}$^{-1}$. It is $\gamma=0$ in the case of a solid body and $\gamma\rightarrow\infty$ in the case of an inviscid fluid. Between these two extremes, we have the viscous bodies, which, under stress relax towards the equilibrium, but not instantaneously. 

We shall mention here the very similar equation used by Darwin in his first paper on the precession of a viscous Earth (Darwin 1877)\footnote{We may paraphrase one of Darwin's statements by just changing the symbols used in it by those shown inside brackets: 
\textit {But because of the Earth's viscosity,} [$\zeta$] \textit{always tends to approach} [$\rho$]. \textit{The stresses introduced in the Earth by the want of coincidence of} [$\zeta$] \textit{with} [$\rho$] \textit{vary as} [$\rho-\zeta$] \textit{. Also the amount of flow of a viscous fluid, in a small interval of time, varies jointly as that interval and the stress. Hence the linear velocity (on the map), with which } [$\zeta$] \textit{approaches} [$\rho$], \textit{varies as} [$\rho-\zeta$]. \textit{Let this velocity be} [$\gamma(\rho-\zeta)$], \textit{where } [$\gamma$] \textit
{depends on the viscosity of the Earth, decreasing as the viscosity increases.}} 
to define its rate of adjustment to a new form of equilibrium, soon extended to the study of tides (Darwin 1879) by means of a law similar to Eq. (\ref{eq:ansatz}). However, in the later paper, he was rather interested in the ocean tides upon a yielding nucleus and was not satisfied with the results that he classed as fallacious. \textit {For unless the viscosity} [of the Earth] \textit{was much larger than that of pitch, the viscous sphere would comport itself sensibly like a perfect fluid, and the ocean tides would be quite insignificant}. This is perhaps the reason for which he never went beyond the circular approximation later used (Darwin 1880) to introduce the tide lag and the tide height in the Earth model in the study of the secular changes of the orbit of the Moon and the changing rotation of the Earth. 

It is also possible to obtain Eq. (\ref{eq:ansatz}) by integrating a spherical approximation of the Navier-Stokes equation of a radial flow across the two surfaces, for very low Reynolds number (Stokes flow), a case in which the inertia terms can be neglected and the stress due to the non-equilibrium may be absorbed into the pressure terms (see Happel and Brenner, 1980). 
The boundary conditions are $\dot{\zeta}=0$ at $\zeta=\rho$.
The pressure due to the body gravitation is given by the weight of the mass which lies above (or is missing below\footnote{This does not mean that a negative mass is being assigned to void spaces; it means just that the forces included in the calculation of the equilibrium figure need to be subtracted when the masses creating them are no longer there.}) 
the equilibrium surface, that is, $-w (\zeta-\rho)$; the modulus of the pressure gradient is the specific weight $w$. 

This comparison allows us to see that the relaxation factor $\gamma$ is related to the viscosity coefficient $\eta$ through
\beq\label{eq:eta}
\gamma=\frac{w R}{2\eta} = \frac{3gm}{8\pi R^2 \eta},
\endeq
where $g$ is the gravity at the surface of the body and $R$ is its mean radius\footnote{Darwin (1879) used a very complete construction of the Navier-Stokes equations and, in his results, the numerical factor is 3/38 instead of 3/8. His numerical factor is determined by the spheroidal form of the tidal potential, but the intensity of the potential does not appear in the result. So, his result would hold even for an infinitesimal tide!}.
This equation is important because it allows us to estimate the range of possible values of $\gamma$ for the celestial bodies to be considered in the applications.

\subsection{The creep equation}

The function $\rho$ may be written as
\beq
\rho=R_e\left(1+\frac{1}{2}\epsilon_\rho \cos 2 \Psi\right)
\endeq
(to the order ${\cal{O}}(\epsilon_\rho)$), where $\Psi$ is the angular distance of one generic point on the surface of the equilibrium ellipsoid to the axis of the tidal bulge (that is, the direction of $\tens{M}$).

\begin{figure}[t]
\centerline{\hbox{
\includegraphics[height=5cm,clip=]{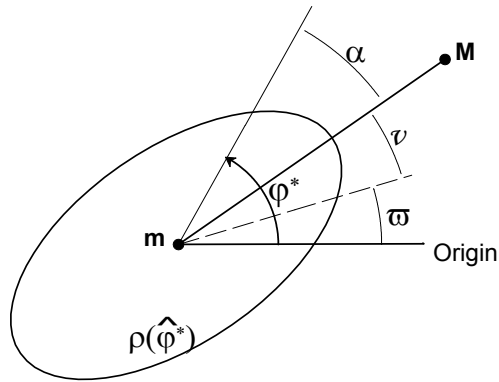}}}
\caption{Equatorial section of the equilibrium spheroid corresponding to the tide generated by $\tens{M}$ on  $\tens{m}$. Angles: $ \alpha$ is the distance from the generic surface point to the vertex of the spheroid; $v$ is the true anomaly of $\tens{M}$; $\varpi$ is the angle between the origin meridian of the body 
and the pericenter of the orbit of $\tens{M}$; $\widehat\varphi^*=\varpi+v+ \alpha$.}
\label{angulos}       
\end{figure}

If we restrict the present study to the case of a ``planar" problem, in which the orbital plane of the tide generating body $\tens{M}$ cuts the body $\tens{m}$ symmetrically (i.e. $\tens{m}$ is symmetrical with respect to an equator, and $\tens{M}$ lies on the same plane as the equator of $\tens{m}$), the differential equation of the adopted model for the creep becomes
\beq
\dot{\zeta}+\gamma \zeta=\gamma\rho = \gamma R^\prime +\frac{1}{2}\gamma R_e \epsilon_{\rho} 
\sin^2{\widehat\theta^*}\cos(2\widehat\varphi^*-2\varpi-2v)
\label{eq:edo}\endeq
where $\widehat\theta^*$ is the co-latitude 
(introduced through $\cos\Psi=\cos\alpha \sin\widehat\theta^*)$,
$R^\prime=R_e(1-\frac{1}{2}\epsilon_\rho\cos^2\widehat\theta^*)$ and 

\beq
\widehat\varphi^*=\varpi+v+ \alpha
\label{eq:varphi}
\endeq
(see fig. 2). 

In the solution of this equation, we have to consider that $\frac{d\widehat\varphi^*}{dt}=\Omega$, angular velocity of rotation of the body $\tens{m}$, which is assumed to rotate in the same direction as the orbital motion of $\tens M$.

\subsection{The circular approximation}\label{circular}
For the sake of making clear which are the main consequences of the proposed rheophysical model, before considering the full model, we consider first the simple case in which the relative motion of the two bodies is circular.
In such case, $r=a$ (semi-major axis) and $v=\ell=nt$ (mean anomaly). 

The resulting equation is a trivial non-homogeneous first-order differential equation with constant coefficients whose solution is
\beq
\zeta={\cal{C}}e^{-\gamma t} + R^\prime + {\cal{A}} \cos (2 \alpha - \sigma_0) 
\label{eq:nonelastic}\endeq
where ${\cal{C}}$ is an integration constant. ${\cal{A}}, \sigma_0$ are undetermined coefficients which may be obtained by simple substitution in the differential equation and identification as\\

\beq
\sigma_0=\arctan \frac{\nu}{\gamma},
\label{eq:ep0}
\endeq

\beq
{\cal{A}}=\frac{1}{2} R_e \epsilon_\rho^\prime \cos\sigma_0
=\frac{\frac{1}{2} R_e \gamma \epsilon_\rho^\prime}{\sqrt{\gamma^2+\nu^2}}
\endeq
where $\epsilon_\rho^\prime=\epsilon_\rho \sin^2\widehat\theta^*$ and  $\nu$ is the semi-diurnal frequency \\

\beq
\nu=2\dot{\alpha}=2\Omega-2n.
\label{eq:nu}\endeq

The integration constant ${\cal{C}}$ depends on $\widehat\varphi^*$ (the integration was done with respect to $t$) and may be related to the initial surface $\zeta_0=\zeta(\widehat\varphi^*,\widehat\theta^*,0)$ through
\beq
{\cal{C}}=\zeta_0-R^\prime - {\cal{A}} \cos (2\alpha(0) - \sigma_0). 
\endeq
The solution depends on $\widehat\theta^*$ via its influence on the constants $R^\prime$ and 
${\cal{A}}$.


\section{Tidal deformation of the body}\label{fullmodel}

In order to develop the theory, we have to consider the two-body equations and introduce
\beq
r=\frac{a(1-e^2)}{1+e\cos v}
\endeq
and 
\beq
v=\ell+(2e - \frac{e^3}{4})\sin\ell+\frac{5e^2}{4}\sin 2\ell+\frac{13e^3}{12}\sin 3\ell+{\cal O}(e^4)
\endeq
into eq. (\ref{eq:edo}). The resulting creep equation is
\beq
\dot{\zeta}+\gamma\zeta =\gamma R^\prime +\frac{15\gamma R_e\sin^2\widehat\theta^*}{8}
\Bigg{(}\frac{M}{m}\Bigg{)}\Bigg{(}\frac{R_e}{a}\Bigg{)}^3 \Bigg{(}
\cos(2\widehat\varphi^*-2\varpi-2\ell )+
\label{eq:edo-e3}\endeq
\begdi
\frac{e}{2}\Big(7\cos(2\widehat\varphi^*-2\varpi-3\ell )-\cos(2\widehat\varphi^*-2\varpi-\ell)\Big)
+\frac{e^2}{2}\Big(-5\cos(2\widehat\varphi^*-2\varpi-2\ell )+17\cos(2\widehat\varphi^*-2\varpi-4\ell )\Big)+
\enddi\begdi
\frac{e^3}{16}\Big(-123\cos(2\widehat\varphi^*-2\varpi-3\ell )+\cos(2\widehat\varphi^*-2\varpi-\ell)+\frac{845}{3}\cos(2\widehat\varphi^*-2\varpi-5\ell )+\frac{1}{3}\cos(2\widehat\varphi^*-2\varpi+\ell)\Big)\Bigg{)}+\cdots
\enddi
or
\beq\label{edo-ultima}
\dot{\zeta}+\gamma\zeta =\gamma R^\prime +\frac{15\gamma R_e\sin^2\widehat\theta^*}{8}\Bigg{(}\frac{M}{m}\Bigg{)}
\Bigg{(}\frac{R_e}{a}\Bigg{)}^3 \sum_{k=-N}^N E_{2,k}(e)\cos(2\widehat\varphi^*-2\varpi+(k-2)\ell)
\endeq
where $N$ is the adopted order of approximation of the Fourier series and the $E_{2,k}(e)$ are the eccentricity functions appearing as coefficients in eq. (\ref{eq:edo-e3}). They are some of the Cayley expansions (Cayley, 1861). An elementary calculation using simple concepts of Fourier analysis shows that 
\beq\label{Fourier-ell}
E_{2,k}(e)=\frac{1}{2\pi}\int_0^{2\pi}\left(\frac{a}{r}\right)^3
\cos\big(2v+(k-2)\ell\big)\ d\ell.
\endeq
(see Appendix \ref{higherorders}).

The integration of eq. (\ref{edo-ultima}) is trivial. 
If we write $\dot{\zeta}+\gamma\zeta = F(t)$, we know that the general solution is 
\beq
\zeta=e^{-\gamma t} \int_t F(t)e^{\gamma t}dt
\endeq
or
\beq
\zeta =C e^{-\gamma t} + R_e + R^{\prime\prime}(\widehat\theta^*, t) +\frac{15\gamma R_e\sin^2\widehat\theta^*}{8}\Bigg{(}\frac{M}{m}\Bigg{)}\Bigg{(}\frac{R_e}{a}\Bigg{)}^3 \sum_{k=-N}^N \frac{E_{2,k}(e) \cos(2 \bar\alpha + k \ell - \sigma_k)}
{\sqrt{\gamma^2+(\nu+kn)^2}}
\endeq
where $\bar{\alpha}=\widehat\varphi^*-\varpi-\ell$ 
(N.B. $\bar{\alpha}-\alpha=v-\ell$),
\begdi
R^{\prime\prime}=-\frac{1}{2} R_e \cos^2\widehat\theta^* e^{-\gamma t} \int_t \epsilon_\rho e^{\gamma t}dt.
\enddi
and
\beq
\sigma_k = \arctan\left(\frac{k n + \nu}{\gamma}\right). \label{eq:sigmas}
\endeq

It is worth emphasizing that the $\sigma_k$ are not {\textit{ad hoc}} lags plugged by hand, but constants introduced during the (exact) integration of the creep equation just to allow us to write the solution in simpler form. 
However the $\sigma_k$ play a role similar to the $\ep_k$ introduced as {\textit {ad hoc}} delays in FRH\footnote{The subscripts used for the $\sigma_k $ are not the same subscripts used in FRH for their homologous lags $\ep_k$.}. 
From their definition, it is also clear that the 
$\sigma_k$ are not small quantities as the lags are assumed to be in standard Darwin theories. 

The solution of the differential equation can also be written as 
\beq\label{eq:zeta-e}
\zeta =C e^{-\gamma t} + R^{\prime\prime} +\frac{1}{2}R_e \sin^2\widehat\theta^* \sum_{k=-N}^N \epsilon_k(e) \cos(2 \bar\alpha + k \ell - \sigma_k),
\endeq
where we have introduced

\beq
\epsilon_k \speq \frac{15}{4}E_{2,k}(e) \cos\sigma_k \Bigg{(}\frac{M}{m}\Bigg{)}
\Bigg{(}\frac{R_e}{a}\Bigg{)}^3 .
\label{eq:prolatenesses}
\endeq
$\zeta$ is formed by the superposition to one sphere of the bulges of several spheroids whose prolatenesses are the $\epsilon_k$.


\section{The attraction of the tidally deformed body}\label{Forces}
In order to proceed, we assume that the potential of $\tens{m}$ is the sum of the potential of one sphere plus the potentials due to various ellipsoid bulges the intersections of which with the equatorial plane are given by the sum of terms in eq. (\ref{eq:zeta-e}). 

\begin{figure}[t]
\centerline{\hbox{
\includegraphics[height=5cm,clip=]{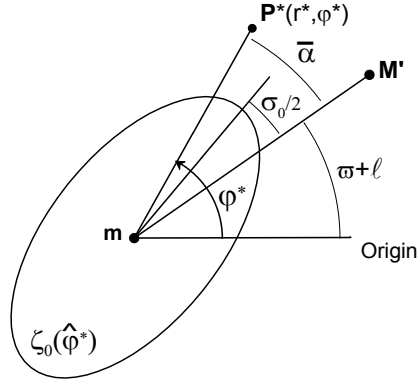}}}
\caption{Main component of $\zeta(\widehat\varphi^*,t)$. Angles: $\bar{\alpha}$ is the distance from the generic point to the direction defined by $\tens{M}'$ (fictitious body in uniform motion; $\tens{M}'\equiv\tens{M}$ when $\tens{M}$ is at the pericenter); $\sigma_0/2$ is the lag of the actual ellipsoid $\zeta_0$ w.r.t. the sub-$\tens{M}'$ point (i.e. the axis of the mean equilibrium ellipsoid). The figure corresponds to the case $\nu > 0$.}
\label{fig:zeta0}       
\end{figure}

For instance, when $\sin\widehat\theta^*=1$, the main component of $\zeta(\bar{\alpha})$ corresponds to 
\beq 
R_e + \frac{1}{2}R_e \epsilon_0(e) {\cos(2 \bar{\alpha} -\sigma_0)},
\endeq
which is the equatorial boundary of a spheroid of mean equatorial radius $R_e$ and prolateness $\epsilon_0$
displaced of an angle ${\sigma_0}/{2}$ with respect to the direction of $\tens{M}'$ (fictitious body in uniform motion; $\tens{M}'\equiv\tens{M}$ when $\tens{M}$ is at pericenter).
(See Fig.\ref{fig:zeta0}). 

The equation of this spheroid is
\beq
\zeta_0 = R_e \left(1 + \frac{1}{2}\epsilon_0 \cos 2\Psi_0\right) = 
R_e \left(1- \frac{1}{2}\epsilon_0 \cos^2\widehat\theta^*+\frac{1}{2}\epsilon_0
\sin^2\widehat\theta^*\cos(2 \bar{\alpha} -\sigma_0)\right).
\label{eq:zeta0}\endeq
where $\Psi_0$ is the angular distance from the generic point to the vertex of the spheroid.  

$\zeta_0$ differs from the component selected in eq. (\ref{eq:zeta-e}) by the additional term $-\frac{1}{2}\epsilon_0 \cos^2\widehat\theta^*$; as discussed in section \ref{sec:axial}, the contribution of this additional term can be neglected in the planar approximation of this study.
 
The disturbing potential (i.e. the potential to be added to the potential of a sphere) due to $\zeta_0$, on an external point $\tens{P}^*(r^*,\theta^*,\varphi^*)$, is 
\beq
\delta U_0=-\frac{2k_fGmR^2\epsilon_0}{15r^{*3}}\big(3\cos^2\Psi_0-1\big)
=-\frac{k_fGmR^2\epsilon_0}{15r^{*3}}\Big(\sin^2\theta^*\big(3\cos(2\bar{\alpha}-\sigma_0)+1\big)-2\cos^2\theta^*\Big)
\endeq
where $G$ is the gravitation constant, $\bar{\alpha}-{\sigma_0}/{2}$ is
the angle between the major axis of the spheroid and the meridian passing by the point $\tens{P}^*$ and 

\beq
k_f=\frac{15A}{4mR^2}
\label{eq:Lovefluid}\endeq
is the fluid Love number. We prefer to use $k_f$ instead of $A$, moment of inertia of $\tens{m}$ with respect to its major axis, because $k_f$ is an adimensional quantity. We remind that for a homogeneous sphere $k_f=1.5$ and that this value decreases as the mass of the body is more concentrated in its central part\footnote{In standard theories using Love numbers, often $k_f$ is substituted by some smaller value to take into account the rigidity of the body. Here, a substitution of this kind is not allowed since $k_f$ is just being used as a convenient substitute for $A$.}.
From now on, since the prolateness appears multiplying all terms, we may use the mean radius $R$ instead of $R_e$. The differences due to this change are of higher order and may be neglected. However, when dealing with the first term of eqn. (\ref{eq:zeta0}), we have to remember that, in the prolate spheroid, $R_e\sim R(1+\epsilon_0/6)$.

To obtain the force acting on one mass located in $\tens{P}^*$, we have to take the negative gradient of $U_0$ and multiply the result by the mass of the point. It is worth emphasizing that $(r^*,\theta^*,\varphi^*)$ are spherical coordinates of one point external to the body while in the preceding equations $(\zeta,\widehat\theta^*,\widehat\varphi^*)$ represented the spherical coordinates of one point on the surface of the body.

Since we are interested in the force acting on $M$ due to the tidal deformation of $\tens{m}$, once the gradient is calculated, we can substitute $(r^*,\theta^*,\varphi^*)$ by the coordinates of $\tens{M}$: $(r,\frac{\pi}{2},\varphi\defeq \varpi+v)$. We remind that this identification  cannot be done before the gradient is computed because, in $\delta U$, we have both, the coordinates $(r,\varphi)$ of the body $\tens{M}$, and the coordinates $(r^*,\varphi^*)$ of the generic point $\tens{P}^*$ (where the gradient is taken).
We thus obtain,
\beq\begin{array}{l@{\speq}l}
F_1&\displaystyle 
-\frac{k_fGMmR^2\epsilon_0}{5r^{4}}\Big(3\cos(2v-2\ell-\sigma_0)+1\Big) \\
F_2&0\\
F_3&\displaystyle
- \frac{2k_fGMmR^2\epsilon_0}{5r^{4}}\sin (2v-2\ell-\sigma_0). 
\end{array}\endeq
The component $F_2$  (force along the meridian) is zero because we are considering here only the planar case. The corresponding torque is
\beq\begin{array}{l@{\speq}l}
M_1&0\\
M_2&\displaystyle
 \frac{2k_fGMmR^2\epsilon_0}{5r^{3}}\sin (2v-2\ell-\sigma_0)\\
M_3&0.
\end{array}\endeq

For the other terms of $\zeta(\widehat\varphi^*)$ we have similar expressions having just to pay attention that, in these terms, $k\ell$ appears added to the arguments.  
We can proceed in the same way as above because the operations done involve only geometric quantities. We thus have 
\beq
\delta U
=\sum_{k=-N}^N -\frac{k_fGmR^2\epsilon_k}{15r^{*3}}\Big(\sin^2\theta^*\big(3\cos(2\bar{\alpha}+k\ell-\sigma_k)+1\big)-2\cos^2\theta^*\Big)
\endeq
where the $\epsilon_k$ are the prolatenesses  defined by Eqn. (\ref{eq:prolatenesses}).
The corresponding force and torque are
\beq\begin{array}{l@{\speq}l}
F_1&\displaystyle \sum_{k=-N}^N
-\frac{k_fGMmR^2\epsilon_k}{5r^{4}}\Big(3\cos\big(2v+(k-2)\ell-\sigma_k\big)+1\Big) \vspace{2mm}\\
F_3&\displaystyle \sum_{k=-N}^N
- \frac{2k_fGMmR^2\epsilon_k}{5r^{4}}\sin \big(2v+(k-2)\ell-\sigma_k\big) \vspace{2mm}\\
M_2&\displaystyle \sum_{k=-N}^N
 \frac{2k_fGMmR^2\epsilon_k}{5r^{3}}\sin \big(2v+(k-2)\ell-\sigma_k\big).
\end{array}\endeq

\subsection{The axial terms}\label{sec:axial}
At last, we have to consider the term $R^{\prime\prime}$, not yet considered, and the terms $\frac{1}{2}R_e\epsilon_k\cos^2\widehat\theta^*$ subtracted from the parts of $\zeta$ to complete the equation of the spheroids. Putting then together, we obtain
\beq
\delta_{\rm ax}\zeta=\frac{1}{2}R_e\Big(\sum_{k=-N}^N \epsilon_k - e^{-\gamma t} \int_t \epsilon_\rho e^{\gamma t} dt \Big) \cos^2\widehat\theta^*.
\endeq
$r=R+\delta_{\rm ax}\zeta$ is the equation of a spheroid with symmetry axis perpendicular to the orbit and whose prolateness is given by the bracket in the above equation. 
As is well-known, the resulting field is axial.
The force on $M$ is central and its contribution to the torque is null.
It will be important in the general non-planar problem because it will contribute for the precession of the axis of the body. Because of its sign, it will counteract the effects due to the oblateness of the body (not considered in the present study).
It will contribute short-period variations in the semi-major axis and eccentricity, which will be averaged to zero over one orbit. Hence, they can be neglected in the planar case.


\section{Rotation of close-in companions}\label{sec:rot} 
We use the equation $C\dot{\Omega}=M_2$ ($\equiv -M_z$) (see FRH \ Sec.7-8).
The time average of  $\dot{\Omega}$ over one period is
\begdi
<\dot{\Omega}>\speq \frac{1}{2\pi C}\int_0^{2 \pi} M_2 \ d\ell. 
\enddi
Hence,  using the approximation $A\simeq C$ in $k_f$ and simplifying:
\beq\label{eq:Omega-av}
<\dot{\Omega}>\speq -\frac{45 G M^2 R^3}{16m a^6 }\left[
\left(1 -  5 e^2  + \frac{63}{8} e^4 - \frac{155}{36} e^6\right) \sin 2 \sigma_0 + 
\right.\hspace*{3cm}\endeq\begdi 
\left(\frac{1}{4} e^2 - \frac{1}{16} e^4 + \frac{13}{768} e^6 \right) \sin 2\sigma_1 +
\left(\frac{49}{4} e^2 - \frac{861}{16} e^4  +  \frac{21975}{256} e^6 \right) \sin 2 \sigma_{-1} + 
\enddi\begdi\left.
\left(\frac{289}{4} e^4 - \frac{1955}{6} e^6 \right) \sin 2\sigma_{-2} 
+ \frac{1}{2304} e^6 \sin 2\sigma_3 + \frac{714025}{2304} e^6 \sin 2\sigma_{-3}\right].
\enddi

It is worth mentioning that only squares will contribute to the average and the above result may be written as
\beq
<\dot{\Omega}>\speq -\frac{45 G M^2 R^3}{16m a^6 }
\sum_{k=-N}^N E_{2,k}^2(e)\sin 2\sigma_k.
\endeq

In order to have an explicit equation in terms of the relaxation parameter $\gamma$ and the involved frequencies, the definitions given by eqs. (\ref{eq:sigmas}) may be introduced into the above equation 
through the trigonometric relation $\sin 2X = 2\tan X / (1+\tan^2 X)$, that is

\beq
\sin 2\sigma_k=\frac{2\gamma(\nu+k n)}{\gamma^2+(\nu+k n)^2}.
\label{eq:senos2}
\endeq

The resulting expression can be used to study the tidal despining of close-in companions and/or central bodies.


\section{Synchronization. Spin-orbit resonance}\label{sec:synchro}
The immediate consequence of eq. (\ref{eq:Omega-av}) is that the synchronous rotation is not a stationary solution of the system when the orbital eccentricity is not zero. Indeed, introducing eq. (\ref{eq:senos2}) and making $\nu=0$, there results, in the first approximation,
\beq
<\dot{\Omega}>\Big|_{\nu=0} \simeq \frac{135 G M^2 R^3n \gamma e^2}{2m a^6 (n^2 + \gamma^2) }.
\endeq
The equality to zero is not possible if $\gamma e\ne 0$. In the synchronous state, the torque is positive, meaning that the rotation is being accelerated by the tidal torque. The stationary solution can only be reached at a supersynchronous rotation. Indeed, solving the equation $<\dot{\Omega}>=0$, we obtain
\beq \label{eq:estacio}
\Omega=n+\frac{6 n \gamma^2}{n^2 + \gamma^2}\ e^2 +
3 n \gamma^2 \frac{226 n^6 + 1453 n^4 \gamma^2 + 
   28 n^2 \gamma^4 + \gamma^6}{8 (n^2 + \gamma^2)^3 (4 n^2 + \gamma^2)}\ e^4 + {\cal O}(e^6).
\endeq

The result corresponds to a supersynchronous rotation. However, at variance with the standard theories, the stationary rotation speed is not independent of the body rheology. 
It depends on the viscosity $\eta$ through the relaxation factor $\gamma$. 

In the quasi-inviscid limit, $\eta\rightarrow 0$, then $\gamma \gg n$ and $\frac{\gamma^2}{\gamma^2+n^2}\simeq 1$. We then obtain 
\beq 
\Omega_{\rm lim}\simeq n (1 + 6e^2 +\frac{3}{8}e^4).
\endeq
Thus, in the quasi-inviscid limit, the result is the same obtained with Darwin's theory when we neglect the differences in the response factors $k_i$ and assume that the \textit {ad hoc} lags of tide components with equal frequencies are equal 
(see Laskar et al. 2004; FRH, sec. 9)


\section{Energy dissipation}
The rate of the work done by the tidal forces is $\dot{W}_{\rm orb}=\mathbf{F}\mathbf{v}$. 

The components of $\mathbf{v}$ in the adopted 3D spherical coordinates are
\beq\begin{array}{l@{\speq}l}
v_1&\displaystyle \frac{nae\sin v}{\sqrt{1-e^2}}\\
v_2&0\\
v_3&\displaystyle\frac{na^2\sqrt{1-e^2}}{r}.
\end{array}\endeq

The result, time-averaged over one period, is:
\beq\label{eq:dotWorb}
<\dot{W}>_{\rm orb}=\frac{3 k_f G M^2 R^5  n}{4 a^6}
\left[ (1  - 5 e^2  +  \frac{63}{8} e^4  - \frac{155}{36} e^6)\sin 2\sigma_0
+ (\frac{1}{8}e^2 - \frac{1}{32} e^4  + \frac{13}{1536} e^6)\sin 2\sigma_1+ \right.
\endeq\begdi\left.
(\frac{147}{8}e^2  - \frac{2583}{32} e^4+ \frac{ 65925}{512} e^6)\sin 2\sigma_{-1}
+ (\frac{289}{2}e^4 - \frac{1955}{3} e^6) \sin 2\sigma_{-2}
-  \frac{e^6}{4608} \sin 2\sigma_3 
+  \frac{3570125e^6}{4608} \sin 2\sigma_{-3}\right]
\enddi
or
\beq
<\dot{W}>_{\rm orb}=\frac{3 k_f G M^2 R^5  n}{8 a^6}
\sum_{k=-N}^N (2-k)E_{2,k}^2(e)\sin 2\sigma_k.
\label{eq:dotWgeral}
\endeq

In the pseudo-synchronous stationary rotation, we may use $\nu$ as given by eq. (\ref{eq:estacio}). Hence, using the $\sin 2\sigma_k$ values given in eq. (\ref{eq:senos2}) 
and  neglecting terms of order higher than ${\cal{O}}(e^2)$, 
\beq\label{eq:dotWstat}
<\dot{W}>_{\rm orb\ (stat)} \simeq -\frac{75 k_f G M^2 R^5 n e^2}{8 a^6}\ \frac{\gamma n}{\gamma^2+n^2}.
\endeq

In addition, we have to consider the work done by the tidal torque on the rotating body: 
$<\dot{W}>_{\rm rot} =C\Omega <\dot{\Omega}>$, that is,
\beq
<\dot{W}>_{\rm rot} \speq -\frac{3 k_f G M^2 R^5 \Omega}{4 a^6 }
\sum_{k=-N}^N E_{2,k}^2(e)\sin 2\sigma_k.
\endeq
(Since $\dot{W}\propto \dot\Omega$, the work associated with the rotation of the body vanishes when it reaches the stationary state.) 

The rate of the mechanical energy released inside the body is
\beq
<\dot{E}>=-\big(<\dot{W}>_{\rm orb}+<\dot{W}>_{\rm rot}\big) > 0.
\endeq


From eq. (\ref{eq:dotWorb}), since $W_{\rm orb}=\displaystyle -\frac{GmM}{2a}$, we obtain $\dot{a}\displaystyle=\frac{2a^2}{GmM}\dot{W}_{\rm orb}$, i.e. the secular variation of the semi-major axis
\beq
<\dot{a}>=\frac{3 k_f  M R^5  n}{4m a^4}
\sum_{k=-N}^N (2-k)E_{2,k}^2(e)\sin 2\sigma_k.
\endeq
The interpretation, neglecting the terms in $e^2$, is easy. If the body is rotating faster than the orbital motion, then $\nu>0$, $\sigma_0>0$, and $\dot{a}>0$. The tide in $\tens{m}$ causes the bodies $\tens{M}$ and $\tens{m}$ to recede one from another. 
Otherwise they are falling one on another.

Two approximations of this formula are useful: 
\begin{description}\item{(i)}
 The free rotating approximation
\beq
<\dot{a}>_{\rm free}\simeq \frac{3k_fMR^5n\gamma\nu}{ma^4(\gamma^2+\nu^2)}
\endeq
and \\
\item{(ii)}  The pseudo-synchronous approximation
\beq
<\dot{a}>_{\rm stat} \simeq - \frac{75 k_f M R^5 ne^2}{4m a^4}\ \frac{\gamma n}{\gamma^2+n^2}.
\endeq
\end{description}

\subsection{The quality factor of standard theories}\label{sec:Q}
The quality factor $Q$ is a parameter originally introduced to characterize damped oscillators. It expresses the quality of the oscillator in keeping free oscillations alive. It is proportional to the proper frequency of the oscillator and vanishes when no elastic force is acting. Its extension to forced oscillations is not done without ambiguities and we do not use it in this theory. 
Nevertheless, the quality factor $Q$ is widely used and, in the applications, we need to know how to express it in terms of the rheophysical parameters used here.
However, it is worth emphasizing that the formulas given in this section are obtained by mere comparison of some equations of this theory with their equivalents in standard Darwin theories and 
are not valid out of the particular conditions in which they were established. 

We recall that, in standard theories, the quality factor  $Q$ and the tidal Love number $k_2$ cannot be separated one from another; in the following equations, $k_2$ is the tidal Love number and $k_f$ is the fluid Love number. 

In standard theories, two different definitions of $Q$ are used, one when the body is free rotating and another when it is trapped in a stationary pseudo-synchronous state.

\subsubsection{Bodies in free rotation}
In classical theory, in this case, we have
\begdi 
<\dot{W}>\simeq \frac{3k_2GM^2R^5n}{2a^6Q}
\enddi
(see FRH eq. 48) where $Q$ is the inverse of the lag of the semi-diurnal tide ($\ep_0$). Comparing to the eccentricity-independent term of eq. (\ref{eq:dotWorb}), we obtain the equivalence formula (valid only for small eccentricities):
\beq\label{eq:Qnu}
Q = \frac{k_2}{k_f}\frac{(\gamma^2+\nu^2)}{\gamma\nu} =\frac{k_2}{k_f}\left[\frac{1}{2}\sin 2\sigma_0\right]^{-1} = \frac{k_2}{k_f}\left(\chi+\frac{1}{\chi}\right)
\endeq
where we have introduced $\chi=\frac{\nu}{\gamma}$ ($\chi$ is the frequency of the semi-diurnal tide in units of $\gamma$).

It is important to note that $Q$ goes to infinity when $\chi$ (or $\nu$) goes to zero. 
This is so also in the standard theories and is just a consequence of the inadequacy of the quality factor $Q$ to measure dissipation. One may note that the dissipation itself, given by eq. (\ref{eq:dotWgeral}), is not singular for $\nu=0$.

\begin{figure}[t]
\centerline{\hbox{
\includegraphics[height=4cm,clip=]{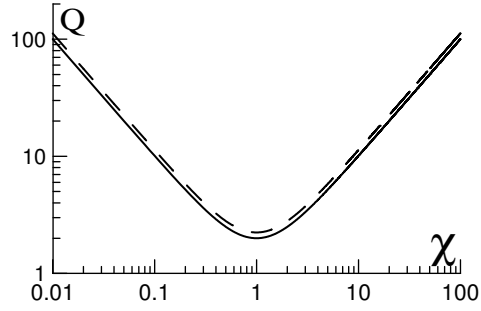}}}
\caption{The low-eccentricity equivalent of the quality factors (in units of $k_2/k_f$) as functions of the frequency (in units of $\gamma$). For rotating bodies, $\chi=\nu/\gamma$ (solid line); for bodies in stationary rotation, $\chi=n/\gamma$ (dashed line).}
\label{fig:Q}       
\end{figure}

\subsubsection{Bodies in pseudo-synchronous stationary rotation}

In standard theories we have, in this case,
\begdi 
<\dot{W}>\simeq -\frac{75k_2GM^2R^5ne^2}{8a^6Q}-\frac{9k_2GM^2R^5ne^2}{8a^6}\ep_5
\enddi
(see FRH eq. 51) where, now, $Q$ is the inverse of the lag of the monthly/annual tide ($\ep_2$). In FRH, we have considered the lag of the radial tide ($\ep_5$) as equal to the lag of the monthly/annual tide ($\ep_2$) since both have the same period in the case of a stationary rotating body. If we compare only the first part of the above equation to eq. (\ref{eq:dotWstat}), we obtain again eq. (\ref{eq:Qnu}). However, when the complete equation is considered, there results
\beq\label{eq:Qstat}
Q_{\rm stat} = \frac{k_2}{k_f}
\frac{(\gamma^2+n^2)}{\gamma n} =\frac{28}{25}\frac{k_2}{k_f}\left[\frac{1}{2}\sin 2\sigma_1\right]^{-1} = \frac{28}{25}\frac{k_2}{k_f}\left(\chi+\frac{1}{\chi}\right)
\endeq
where now $\chi=\frac{n}{\gamma}$ ($\chi$ is the frequency of the monthly/annual tide in units of 
$\gamma$)

This duality in the actually used definitions of $Q$ in standard theories is a big nuisance.
In the case of planetary satellites, eccentricities are low and either the system is rotating or nearly synchronous and we may consider the two cases separately. 
However, in the case of exoplanets, eccentricities are often high and the choice of one of the two formulas to determine $Q$ is a problem.
Indeed, in the standard approach, if the eccentricity is large, the stationary rotation may have a period much smaller than the orbital period and the dissipation due to the semi-diurnal tide will not vanish as in true synchronous companions.
As a consequence, tidal components with frequencies $\nu$ and $n$ will contribute to the energy dissipation on the same foot making impossible to privilege one of them to define one quality factor.

\section{Circularization}

The variation  of the remaining elements can be obtained straightforwardly using Gauss equations (see Beutler 2005, Sec. 6.3.5). As discussed in FRH (section 18.1), in order to take into account correctly the reaction on $\tens{M}$ of its tidal action on $\tens{m}$, the accelerations $R',S',W'$ of those equations need to be multiplied by $(M+m)/m$ or, equivalently, by $n^2a^3/Gm$. With the forces calculated in Section \ref{Forces}, we thus get 
\beq
<\dot{e}>=-\frac{3k_f MR^5ne}{8ma^5} \left[
\big(1 - \frac{21}{4} e^2 +  9 e^4  - \frac{3299}{576} e^6 \big)\sin 2\sigma_0 
+ \big(\frac{1}{4} - \frac{1}{16}e^2 - \frac{35}{768} e^4 - \frac{175}{18432} e^6 \big) \sin 2\sigma_1 + \right.
 \endeq\begdi
\left. \big(-\frac{49}{4} + \frac{1253}{16} e^2  - 
 \frac{50311}{256} e^4 + \frac {508651}{2048} e^6 \big) \sin 2\sigma_{-1} +
\big(- \frac{289}{2} e^2 + \frac{10421}{12} e^4  - \frac{310463}{144} e^6 \big) \sin 2\sigma_{-2} + \right.
\enddi\begdi\left.
\big(\frac{1}{768} e^4 + \frac {17}{18432} e^6 \big) \sin 2\sigma_3 -
\big(\frac{714025}{768} e^4 - \frac {105299675}{18432} e^6 \big) \sin 2\sigma_{-3} +
 \frac{e^6}{144} \sin 2\sigma_4 - \frac{284089e^6}{64} \sin 2\sigma_{-4} 
\right] \enddi
or
\beq\label{eq:doteav}
<\dot{e}>=-\frac{3k_f MR^5n}{8ma^5e}\sum_{k=-N}^N \Big(2\sqrt{1 - e^2} - (2-k)(1 - e^2)\Big)E_{2,k}^2(e)\sin 2\sigma_k.
\endeq
Taking into account eqs. (\ref{eq:senos2}), we may also write
\beq
<\dot{e}>=-\frac{3k_f MR^5ne\gamma}{16ma^4} \left(
4 \frac{\nu}{\gamma^2+\nu^2} - 49 \frac{(\nu-n)}{\gamma^2+(\nu-n)^2} + \frac{(\nu+n)}{\gamma^2+(\nu+n)^2} \right) + {\cal O}(e^3).
\endeq

In the case of a stationary or near-stationary rotation, $\nu={\cal O}(e^2)$ and the above equation is reduced to
\beq
<\dot{e}> \simeq -\frac{75}{8}\frac{k_f MR^5ne\gamma}{ ma^5}\frac{n}{\gamma^2+n^2}.
\endeq

\section{Dissipation parameters in stars, planets and satellites}\label{examples}

In this section we determine the values of $\gamma$ for several Solar System and extrasolar bodies.
We use for that sake the values published in the literature usually obtained using standard tidal evolution theories. One problem common to most of the given examples is that the inversion of the equivalence formulas relating $\gamma$ to $Q$ has two solutions. The choice of one of the two solutions is done after comparing the values of the equivalent uniform viscosity in each solution.  

\subsection{Io}
The tidal evolutions of the Galilean satellites of Jupiter are among the best studied in our Solar System. 
From the satellites accelerations, Lainey et al. (2009) have determined the dissipations of Io and Jupiter. For Io, they have found $k_2/Q=0.015 \pm 0.003$.
Introducing this value in the formulas given in section \ref{sec:Q}, we obtain
$\gamma=4.9 \pm 1.0 \times 10^{-7} $ Hz.
It is worth mentioning that this result is independent of the individual values of $k_2$ and $Q$, which are not well known.  
The calculated value of $\gamma$ depends only on the value of $k_2/Q$ and on the moment of inertia (0.378 $mR^2$). 

With the results given in Sec. \ref{sec:synchro} (eq. \ref{eq:estacio}), we obtain for the synodic rotation period (\textit{a.k.a.} length of the day) $P_{\rm syn}= 3300^{+1800}_{-1000}$ yr.
We may compare this value with the minimum value 1400 yr determined by Milazzo et al. (2001) from the  comparison of \textit{Galileo} and \textit{Voyager} images taken 17 yrs apart.

The equivalent viscosity corresponding to this determination may be obtained using eq. (\ref{eq:eta}). The result, $1.2 \pm 0.3 \times 10^{16}$ Pa s, is in good agreement with the value used by Segatz et al. (1988): $2 \times 10^{16}$ Pa s, in their models for Io's tidal dissipation. 

We note that the adopted value of $\gamma$ corresponds to $\chi=94$, that is, to the ascending branch of the curve shown in fig. \ref{fig:Q} (Darwin's theory corresponds to the descending branch). Therefore, we are in the regime proposed by Efroimsky and Lainey (2007) in which $Q$ increases with the frequency of the tide component. The solution corresponding to Darwin's regime gives $\gamma=0.003$ and $\eta=1.7 \times 10^{12}$, which is $2-3$ orders of magnitude below the known viscosity of ice and silicates (resp $1.5\times 10^{14}$ Pa s and $3\times 10^{15}$ Pa s \textit{cf.} Sotin et al. 2004).

\subsection{Europa}
The basic information we have on Europa is its non-synchronous rotation.
The synodic period is constrained by the lower limit 12,000 yr obtained from the comparison of \textit{Galileo} and \textit{Voyager} images, and the upper limit 250,000 yr, obtained by comparing the present position of some cycloidal cracks with the longitudes at which their shapes should have been formed (see Greenberg et al. 2002). Using the results given in section \ref{sec:synchro}, we obtain $\gamma= 1.8 - 8.0 \times 10^{-7} $ Hz and $k_2/Q$ between 0.01 and 0.045
(if we adopt $k_2=0.26$, we obtain $Q$ in the range  6--26).
The equivalent uniform viscosity corresponding to these results is $\eta=4-18 \times 10^{15}$ Pa s.

The same indetermination discussed above occurs here since the excess of rotation speed is approx. proportional to $\chi^2+\chi^{-2}$. As before, we have to look to the viscosity to decide between the two mathematically possible solutions. 
The solution corresponding to Darwin's regime gives $\gamma=0.0005 - 0.002$ and $\eta=1.3-6 \times 10^{12}$ Pa.s, which is less than expected, but not so expressively as in other cases.

We note that the indetermination is as more difficult to solve as $Q$ is small.

\subsection{The Moon}
The quality factor of the Moon has been determined for two tidal frequencies. For the monthly tide, $Q=30\pm 4$ and for the annual tide $Q \sim 35$ (Williams et al., 2005; Williams and Boggs, 2008).
The values of $Q$ for the monthly tide (combined with the very low $k_2$ of the Moon, 0.0301 \textit{cf.} Williams et al.) gives $\gamma=2.0 \pm 0.3 \times 10^{-9} $ Hz. 
The synodic period of the Moon corresponding to the above determinations of $\gamma$ is larger than 5 Myr and indistinguishable from a true synchronous spin-orbit resonance.

The problem with this determination is the value found for the uniform viscosity: 
$\eta=2.3 \pm 0.3 \times 10^{18}$ Pa s, $4 - 5$ orders of magnitude smaller than the values reported in the literature, which refer to the solid lithosphere. 
This seems to be one more  indication in favor of the role played by a plastic lunar asthenosphere in tidal dissipation and is in agreement with the dramatic decrease of the seismic $Q$ below the deep moonquake source region indicating the presence of a partial melt below this depth and that, likely, most of the solid body dissipation in the Moon occurs below ~1150 km (see Wieczorek et al. 2006). The alternative solution, corresponding to Darwin's regime, gives viscosity values yet smaller. 
The adoption of a more realistic value for the viscosity leads to a much smaller $\gamma$ and a much larger synodic period, even when a permanent equatorial asymmetry of the Moon is neglected. The resulting $\gamma$ is small enough to compensate the fact that $n$ and $e$ may have been larger in the past and to give a result consistent with the rotation indicated by the distribution of the craters on the Moon: a synchronous attitude lasting since the formation of the last great basin, 3.8 Gyrs ago (see Wieczorek and Le Feuvre, 2009). However, a larger viscosity would also imply a larger $Q$, in contradiction with the above-mentioned determinations.\footnote{In the Efroimsky-Layney regime, $Q$ increases with the relative frequency $\chi$ of the tide component and then increases when the critical frequency $\gamma$ decreases.}

Let us add that the annual tide cannot be studied with the Keplerian model used in this paper. It is clearly related to the motion of the Earth-Moon system around the Sun and a perturbed model is necessary to interpret it. 

\subsection{Titan}

Cassini radar observations of Titan over several years show that the present-day rotation period of Titan is different from synchronous and correspond to a shift of $\sim 0.12^\circ$ per year in apparent longitude (Stiles et al. 2008, 2010). This result, if only due to tidal torques, would mean a very large dissipation ($Q < 10$). Values of $Q$ are not reported in recent literature, however, from internal structure studies, Tobie et al. (2005) conclude that because of convection in the outer ice layer, we have now $Q \sim 50$ (but $Q > 300$ during almost the whole satellite past evolution). Adopting this value
and $k_2$ in the interval 0.32--0.39 (\textit{cf} Sohl et al. 2003), we obtain $\gamma = 2.9 \pm 0.2 \times 10^{-8}$ Hz, $\eta= 1.1 \pm 0.1 \times 10^{17}$ Pa s and a synodic period $P \sim 200,000$ years. 
As the Moon, Titan behaves much like a solid body. This favors the interpretation of the measured shift of Titan's crust as due to seasonal effects (e.g. the interaction between the crust and the atmosphere; see Tokano and Neubauer 2005; van Hoolst et al., 2008). 

Let us also mention that a re-analysis of the Cassini data by Meriggiola and Iess (Meriggiola, 2012) has not showed discrepancy from a synchronous motion larger than $0.02^\circ$ per year. 
This more conservative result does not disagree with those of the analysis done above. 
It only sets less strict limits as $\gamma  < 1 \times 10^{-7}$ Hz and $Q > 15$.

\subsection{Jupiter}
From the acceleration of the Galilean satellites, mainly Io, the dissipation in Jupiter has been determined to be given by $k_2/Q=1.102 \pm 0.203 \times 10^{-5}$ (Lainey et al. 2009). 
This value corresponds to $\chi=1.1 \pm 0.2 \times 10^{-5}$ and $\gamma=23 \pm 4 $ Hz. 
The equivalent uniform viscosity is $\eta_{\rm }=4.7 \pm 0.9 \times 10^{10}$ Pa s. Jupiter's tides are in Darwin's regime (the other alternative would need a viscosity as high as  
$10^{20}$ Pa s)

\subsection{Saturn}
From the maximum possible past evolution of Mimas, Meyer and Wisdom (2007) obtained for Saturn, the limit $Q>18,000$. If we use Saturn's Love number, $k_2=0.341$, we obtain $\chi < 2.4 \times 10^{-5}$ (Darwin's regime)  and $\gamma>13$ Hz. The equivalent uniform viscosity is $\eta_{\rm }< 1.5 \times 10^{10}$ Pa s. 

\subsection{Neptune}
Founded on previous studies of the Neptunian satellites, Hamilton (2009) proposes the value $Q/k_2=4.5 \times 10^{4}$ (with an error factor 2). If we adopt the Love number, $k_2=0.41$ (Durda, 1992), there follows $\gamma=9.4$ Hz and $\eta_{\rm }= 2.4 \times 10^{10}$ Pa s. The factor 2 of incertitude in $Q$ is reproduced in the values of $\gamma$ and $\eta$.

\subsection{Solid Earth}
The tidal dissipation in the solid Earth was estimated from satellite tracking and altimetry (Ray et al. 1996) as $Q=370$ with an error factor $\sim 2$ (lag angle $0.16 \pm 0.09$ degrees). Considering the Love number $k_2=0.46$, there follows $\chi=770$, $\gamma=1.8 \times 10^{-7}$ Hz and $\eta=9\times 10^{17}$ Pa s. All these values have error factors $\sim 2$. The viscosity found is several orders of magnitude smaller than the viscosity of the Earth's lower mantle (see Karato, 2008). This uniform value does not make a distinction between the various parts of the solid Earth and is expected to be smaller than the viscosity of the mantle. The result, however, allows us to solve the two-solution indetermination in the inversion of $Q(\chi)$. For instance, in the present case, the root corresponding to a Darwin's tide regime ($\gamma \sim 0.1$ Hz) leads to viscosity values 6 orders of magnitude smaller than the above given one and can be excluded. 
The low viscosity found is an indication that the given $\gamma$ is too large and that we should look for a creeping law leading to more dissipation in high frequencies, as it happens when the Andrade model is plugged in the modified standard theory as proposed by Efroimsky (2012).

\begin{table}[t]
\label{tab:summary}
\caption{Summary of the values adopted and/or obtained in this section. }
\begin{tabular}{lcccc}
 Body & $\gamma$ (Hz) & $2\pi/\gamma$      &$\eta$ (Pa s)  & $Q$ equivalent \\
\hline\\
Moon         & $2.0 \pm 0.3 \times 10^{-9}$&  36,000 d      & $2.3 \pm 0.3 \times 10^{18}$& $30 \pm 4$\\
Titan     & $2.9 \pm 0.2 \times 10^{-8}$& 2500 d      & $1.1\pm 0.1 \times 10^{17}$ & $\sim$ 50   \\
Solid Earth & $0.9-3.6 \times 10^{-7}$&  200-800  d      & $4.5-18 \times 10^{17}$ & 200--800\\
Io       & $4.9\pm 1.0\times 10^{-7}$ &   730 d    & $1.2 \pm 0.3 \times 10^{16} $ & 50 -- 80\\   
Europa & $1.8-8.0\times 10^{-7}$ & 90--400 d         & $4-18 \times 10^{15}$ & 6 -- 26 \\
Neptune &  2.7--19 & $<$ 2 s & $1.2 - 4.8  \times 10^{10}$ & 9,000-37,000\\
Saturn   & $> 7.2 $  &   $< 0.9$ s& $< 15 \times 10^{10}$ & $> 18,000$ \\
Jupiter & $23 \pm 4$   &  $\sim$ 0.3  s  & $4.7 \pm 0.9  \times 10^{10}$ & $\sim$ 36,000  \\
hot Jupiters  & 8--50 & 0.1--0.8 s & $5 \times 10^{10}- 10^{12}$ & $2\times 10^5 - 2\times 10^6$\\
solar-type stars &  $>30$   & $< 0.2$ s &  $< 2 \times 10^{12}$ &  $ >2 \times 10^6 $ \\
\hline 
\end{tabular}
\end{table}

\subsection{Hot Jupiters}\label{sub:hot}

A first assessment of the values of $\gamma$ for hot Jupiter may be done using the values found by Hansen (2010) from an analysis of the survival of some short-period exoplanets of mass $\sim 0.5 M_{\rm Jup}$.
The comparison of our results to the formulas used by Hansen adapted to the case of a planet trapped in a stationary rotation (i.e. pseudo-synchronous), gives
\beq
\gamma=\frac{25 G k_f}{42 R^5 \sigma_p}
\endeq
where $\sigma_p$ is the planetary dissipation parameter used by Hansen. 

Using his mean results for WASP-17 b, Corot-5 b and Kepler-6 b, transiting planets whose radii have been determined, we obtain values $\gamma$ in the range 8 - 50 Hz, the smaller value corresponding to the bloated WASP-17 b (radius $\sim 2R_{\rm Jup}$). 
The results are also sensitive to the moment of inertia of the planets (via $k_f$) and were obtained using $A>0.1 mR^2$. 
If the central concentration is yet larger (i.e. if $A$ is smaller), $\gamma$ may be smaller.

Another possibility is to use the known value of Jupiter's $Q$ and some scaling laws
: (\textbf{i}) If $n\ll \gamma$, $Q$ scales with the period of the main tidal component (see section \ref{sec:Q}); (\textbf{ii}) $Q$ scales with $R^{-5}$ (see Eggleton et al. 1998; Ogilvie and Lin, 2004, Hansen, 2010). 
For instance, we may consider one hot Jupiter with a mass of 2--3 M$_{\rm Jup}$ in a 5-day orbit. The period of the main tide raised on it is 29.4 times the period of the semi-diurnal tide of Jupiter (raised by Io) and planets of this mass range have radii $1.2 \pm 0.2 R_{\rm Jup}$. With these data we obtain $Q \sim 420,000$ and, then,  $\gamma = 15$ Hz. 

The very existence of hot Jupiters around old stars in significantly non-circular orbits is an important test for tidal theories. 
Indeed, all general mechanisms responsible for important eccentricity enhancement are related to events expected to occur in the early stages of the formation of the system (see Malmberg and Davies, 2009).
Therefore, in older systems, if eccentricities were not damped to zero, the variations due to the tidal evolution may have been below some limits. 
We note that the existence of several planets in these conditions\footnote{e.g. CoRot-5 b ($M=0.47 M_{\rm Jup}, e=0.09^{+0.09}_{-0.04})$, CoRoT-12 b ($M=0.9 M_{\rm Jup}, e=0.07^{+0.06}_{-0.04}$) and CoRoT-23 b ($M=2.8 M_{\rm Jup}, e=0.16 \pm 0.07$)}  plays against hypothesizing  that exceptional sources of significant enhancement may have existed in the recent story of each one of them. However, we emphasize that the two parameters considered in this analysis, age and eccentricity, are of difficult determination.

We have studied some of the CoRoT hot Jupiters in elliptic orbit. On one hand, transiting planets have better determined eccentricities\footnote{The true longitude must be 90 degrees at the minimum light of the transit. This constraint added to the radial velocity measurements, allows a better determination of the eccentricity and the longitude of the pericenter than in the case of non-transiting planets where these parameters are to be determined based on hard-to-measure asymmetries of the radial velocity curve.}, and, on the other hand, their ages were determined with some confidence. 

\begin{figure}[t]
\centerline{\hbox{
\includegraphics[height=6cm,clip=]{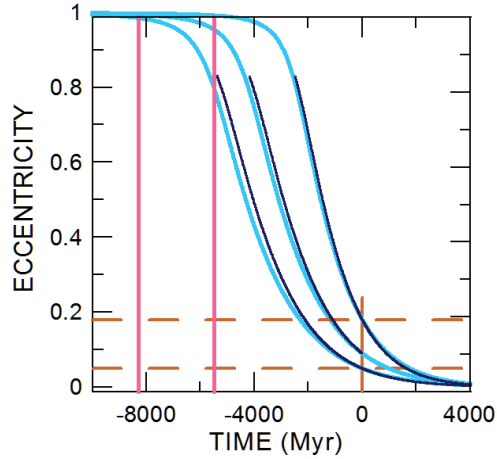}}}
\caption{Simulation of the tidal evolution of the orbit of planet CoRoT 5b using 
$\gamma=200$ Hz (black) for initial eccentricities 0.05, 0.09 and 0.18. 
The corresponding results with the Darwin-Mignard approach using $Q=3.4\times 10^6$ are also shown (blue). 
Vertical lines show $t=0$ and the age of the star range (5.5--8.3 Gyr). 
Horizontal lines: Observed eccentricity at $t=0$ (limits of the error bar \textit{cf.} Rauer et al. 2009).}
\label{fig:simu}       
\end{figure}

Fig. \ref{fig:simu} shows the tidal evolution of the hot Jupiter CoRot-5 b calculated using the approach developed in this paper (black lines) and the standard Darwin theory (blue lines), respectively. 
In order to avoid well-known truncation errors associated with expansions, 
we have used $N$=150 in the expansions, which allows dealing with eccentricities in the range of the solutions shown ($e < 0.85$) without serious truncation errors (see the Appendix). 
The evolution predicted by Darwin's theory was simulated by numerical integration of one 2-body model using  Mignard's expression for the tidal force in closed form.  

Fig. \ref{fig:simu} shows the good agreement of the rheophysical theory of this paper and Darwin's standard theory in the study of the evolution of giant planets. 
The solutions shown in figure \ref{fig:simu} correspond to $\gamma = $ 200 Hz and its equivalent $Q=3.4\times 10^6$, respectively.

\subsection{Hot super-Earths}

The few known transiting planets in the 1-10 Earth mass range are in circular orbits and the memory of their past evolution is erased. 
There is only one case among the currently known ones that may give us some information. 
It is 55 Cnc $e$ for which $e=0.057_{-0.041}^{+0.064}$. It belongs to a somewhat hierarchized system of 5 planets whose past evolution may be simulated from 
\textit{ad hoc} initial conditions able to bring the system to its present situation. 

Simulations with the standard theory (Mignard's torque) starting with arbitrary excited eccentricities showed that dissipation in 55 Cnc e drives the innermost planets (55 Cnc e and 55 Cnc b) to a stationary solution with aligned pericenters (secular mode I of Michtchenko and Ferraz-Mello, 2001) and the eccentricity of 55 Cnc e falls to $\sim 0.003$ in a time much shorter than the age of the system (which is $10.2 \pm 2.5$ Gyr). 
If $Q > 5500$ (dissipation factor of the current ``annual" tide), the damping to the equilibrium center of the secular dynamics is much slower allowing 55 Cnc e to be found at a larger eccentricity now.

The only other possible guess comes from the viscosity of CoRoT-7 b estimated by Leger et al. (2011): $\eta > 10^{18}$ Pa s. Using the physical data of that planet, we obtain $\gamma<5\times 10^{-7}$ Hz and $Q=100$.
This result means that for CoRoT-7 b, $\chi > 1$ and $Q$ grows with the frequency (Efroimsky-Lainey regime), one fact that should be taken into account in the simulations of the evolution of the system. 
If $k_2=0.46$ as for the Earth, then $Q^\prime=3Q/2k_2=300$, which is of the same order as the value 100 adopted by Rodriguez et al. (2011) in the study of the tidal evolution of the CoRoT-7 planets.

\subsection{Solar type stars}

Dissipation values of solar type stars have been estimated by  Hansen (2010) from an analysis of the survival of some short-period exoplanets. 
The comparison of the equations used in Hansen's evolution model to ours allows us to write:
\beq
\gamma=\frac{2 G k_{f \rm star}}{ 3R_{\rm star}^5 \sigma_{\rm star}}
\endeq
where $\sigma_{\rm star}$ is the stellar dissipation parameter used by Hansen. 
Using his mean result $\sigma_{\rm star}= 8.3 \times 10^{-64}$ g$^{-1}$ cm$^{-2}$ s$^{-1}$, we obtain for M and G stars, $\gamma \sim 3 - 25 $ Hz. This variation is mainly due to important dependence on the radius and the upper limit correspond to stars having half of the radius of the Sun. For one star equal to the Sun, the result is $\gamma = 7$ Hz. 

An approximated calculation gives for the corresponding kinematic viscosity, $10^9 - 10^{10}$m$^2$s$^{-1}$,  which are much larger than the value  $10^8 $ m$^2$s$^{-1}$ used by Ogilvie and Lin (2007) in their models of tidal dissipation in stars. 

However, in the study of transiting hot Jupiters mentioned in Sec. \ref{sub:hot}, we have found that $\gamma < 30$ Hz often implies a too large exchange of angular momentum between the orbit and the star rotation.
As a consequence, we needed to assume extremely low values for the star rotation in the past to be able to reproduce current observed values.

The limit $\gamma > 30$ Hz given in table 1 is more or less the same obtained by Jackson et al. (2011) from the analysis of the distribution of the putative remaining lifetime of hot Jupiters. 
 We note that the corresponding limit for the viscosity is 30 times larger than that adopted by Ogilvie and Lin (2007).

\begin{figure}[t]
\centerline{\hbox{
\includegraphics[height=6cm,clip=]{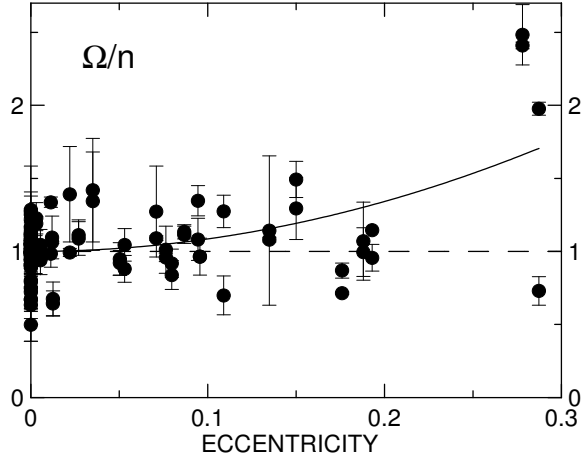}}}
\caption{Binary stars: Distribution of the angular velocity of rotation in function of the eccentricity.}
\label{fig:binary}       
\end{figure}

\subsection{Binary stars}
We may use a sample of data on detached binary stars selected from those collected by Torres et al. (2010), to investigate their rotations. If we fit a law $\lambda e^2$ through the points shown in fig. \ref{fig:binary}, we obtain a coefficient close to 8. However, a further analysis shows that this value is strongly determined by the  few points corresponding to $e>0.2$; when these points are not included the coefficients falls to 2. 
The concentration of the measured values around the synchronous rotation is clearly seen; we may also see that among the stars in elliptic orbits a majority shows rotation above the synchronous value (i.e. they are super-synchronous). However, the high dispersion of the points does not allow us to fix the value of $\lambda$, nor to discard the coefficient corresponding to a standard solution ($\lambda=6$), which may prevail because of the low viscosity (and consequently high $\gamma$) of normal stars.


\section{The elastic tide}\label{sec:lag}

The shape of the body deformation due to the creep tide is given by eqn. (\ref{eq:zeta-e}). After the transient phase (i.e. for $\gamma t \gg 1$), only the forced terms matter and it is dominated by the semi-diurnal component 
\beq\label{eq:semidi}
\delta\zeta = 
\frac{15 R_e\sin^2\widehat\theta^*}{8}\Bigg{(}\frac{M}{m}\Bigg{)}\Bigg{(}\frac{R_e}{a}\Bigg{)}^3 E_{2,0}(e) \cos \sigma_0 \cos(2 \bar\alpha - \sigma_0),
\endeq
the maximum of which is reached when $2 \bar\alpha - \sigma_0 = 0$, i.e. the angle between the vertex of this component to the sub-$\tens{M}^{\prime}$ point is $\sigma_0/2$. We remind that $\sigma_0$ is a constant determined by the integration of the creep equation and not an \textit{ad hoc} lag; it is not necessarily small. In addition, $\tan \sigma_0$ is proportional to the semi-diurnal frequency $\nu$ and inversely proportional to the relaxation factor $\gamma$ (i.e. it is proportional to the viscosity). 

In the case of an inviscid body, the response to the tidal action is instantaneous, $\gamma\rightarrow\infty$ and $\sigma_0\rightarrow 0$; the tide highest point remains aligned with the mean direction of the tide raising body $M$ (as shown in Fig. \ref{angulos}). 

However, when $\gamma\ll\nu$, as in rocky planets and satellites, $\sigma_0$ will approach $90^\circ$ (in the rigid body limit, $\gamma=0$ and $\sigma_0=90^\circ$). 
This result is in contradiction with the observations. For instance, the observed geodetic lag of the Earth's body semi-diurnal tide is very small ($0.16 \pm 0.09$ degrees $cf.$ Ray et al. 1996). In addition, some authors (see Efroimsky, 2012) claim that in these bodies
the actual lags do not obey the weak friction approximation where lag tangents are proportional to the frequencies, but  have lags proportional to a negative power of the frequency.

In order to conciliate the theory and the observed tidal bulges in the Earth, we have to assume that the actual tide is not restricted to the component due to the creeping of the body under the tidal action, but has also a pure elastic component. 
No matter how empirical this hypothesis seems to be, it explains well the observed behavior. 

Let this elastic component be defined at each point by its height over the sphere and be given by 
$\delta\zeta_{\rm el}(\widehat\phi^*,\widehat\theta^*)= \lambda (\rho(\widehat\phi^*,\widehat\theta^*)-R^\prime)$ where $\rho$ is the radius vector of the equilibrium spheroid surface and   
$\lambda$ is a quantity related to the maximum height of the tide (see Sec.\ref{sec:height}).  
For the Earth, for instance, $\lambda \sim 0.2$, which is the ratio of the observed maximum height of the lunar tide (26 cm after Melchior, 1983) to the maximum height of the equilibrium spheroidal figure (1.34m).

The sum of the (local) heights of the elastic tide and of the main term of the creep tide is 
\beq
\delta\zeta=\frac{1}{2}R_e \epsilon_\rho \Big(\lambda \cos 2\alpha +
\cos\sigma_0 \cos(2\bar\alpha-\sigma_0)\Big)
\label{eq:geodetic}
\endeq
where, 
for the sake of simplicity, we have set $E_{2,0}(e)=1$ and $ \sin \widehat\theta^*=1$ (equator). 

There is some similarity between this composite model and the model studied by Remus et al. (2012) and due to Zahn (1966). The elastic tide of this paper is, in principle, the same as Zahn's adiabatic tide. However, the creeping tide is very different from Zahn's dissipative tide. The physical setting of the two models is not the same and the results are different; for instance, the creeping tide is not in quadrature with the exciting potential, as Zahn's dissipative tide.

\begin{figure}[t]
\centerline{\hbox{
\includegraphics[height=5cm,clip=]{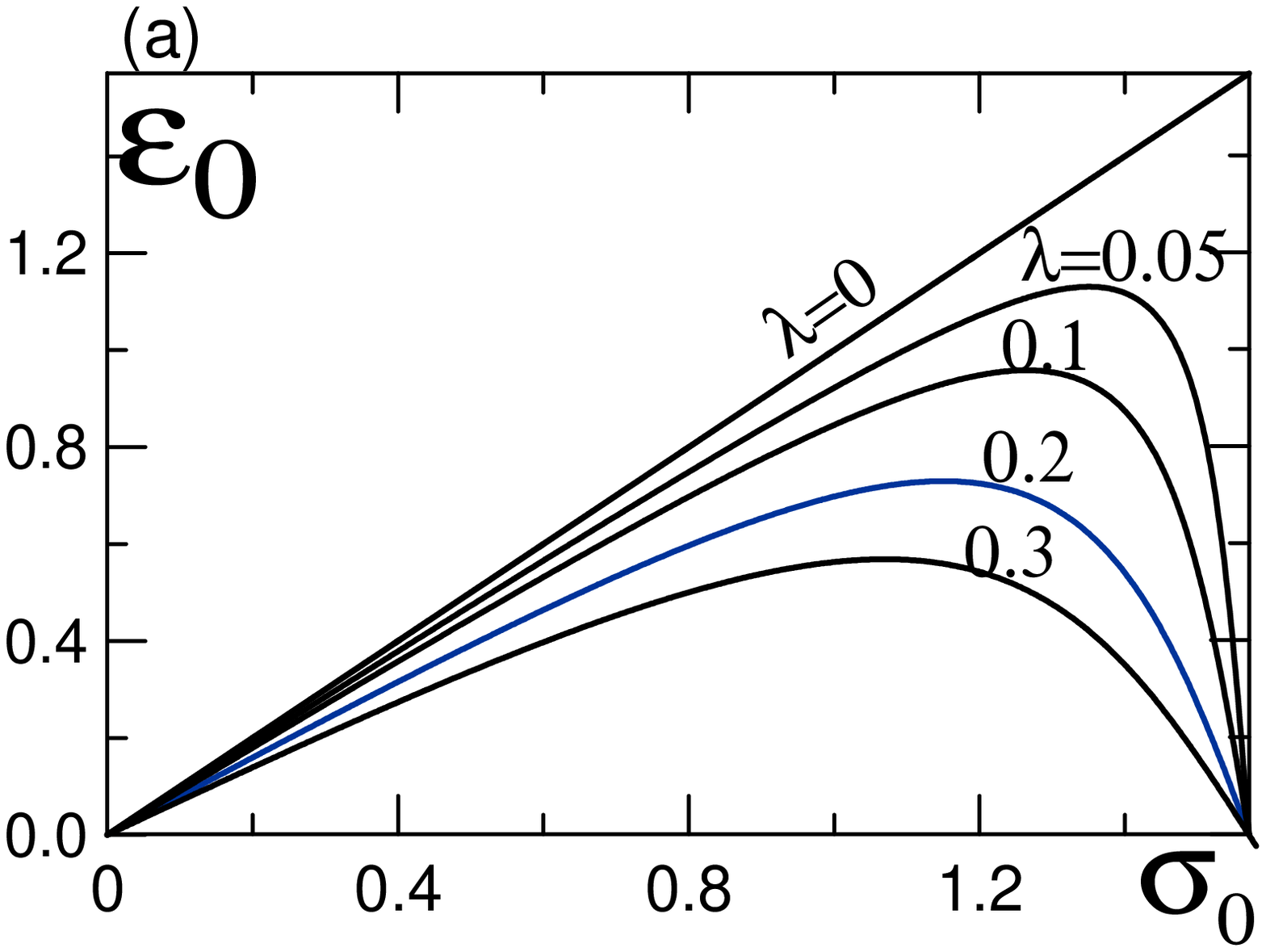}\hspace{3mm}
\includegraphics[height=5cm,clip=]{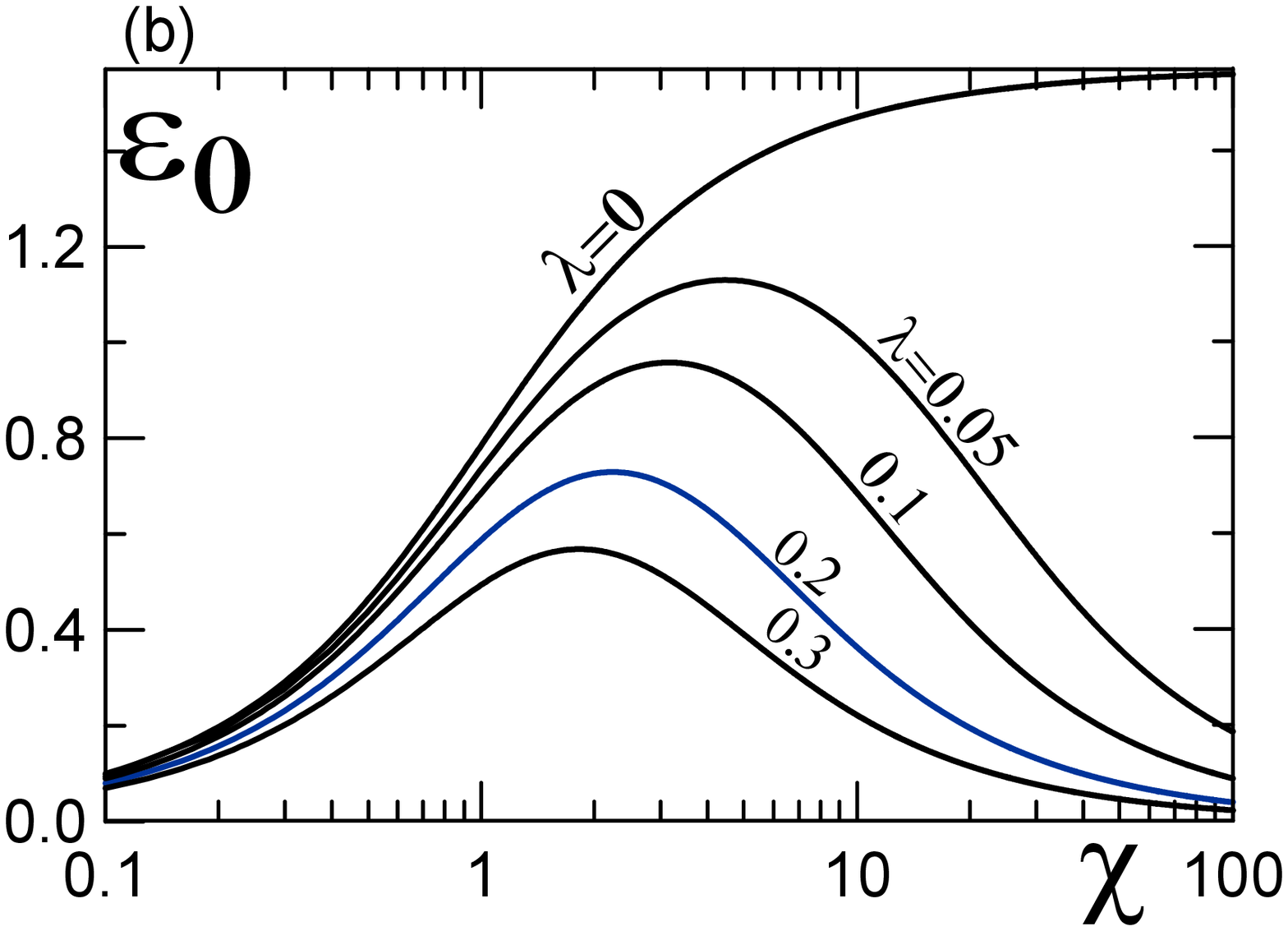}
}}
\centerline{\hbox{
\includegraphics[height=4.7cm,clip=]{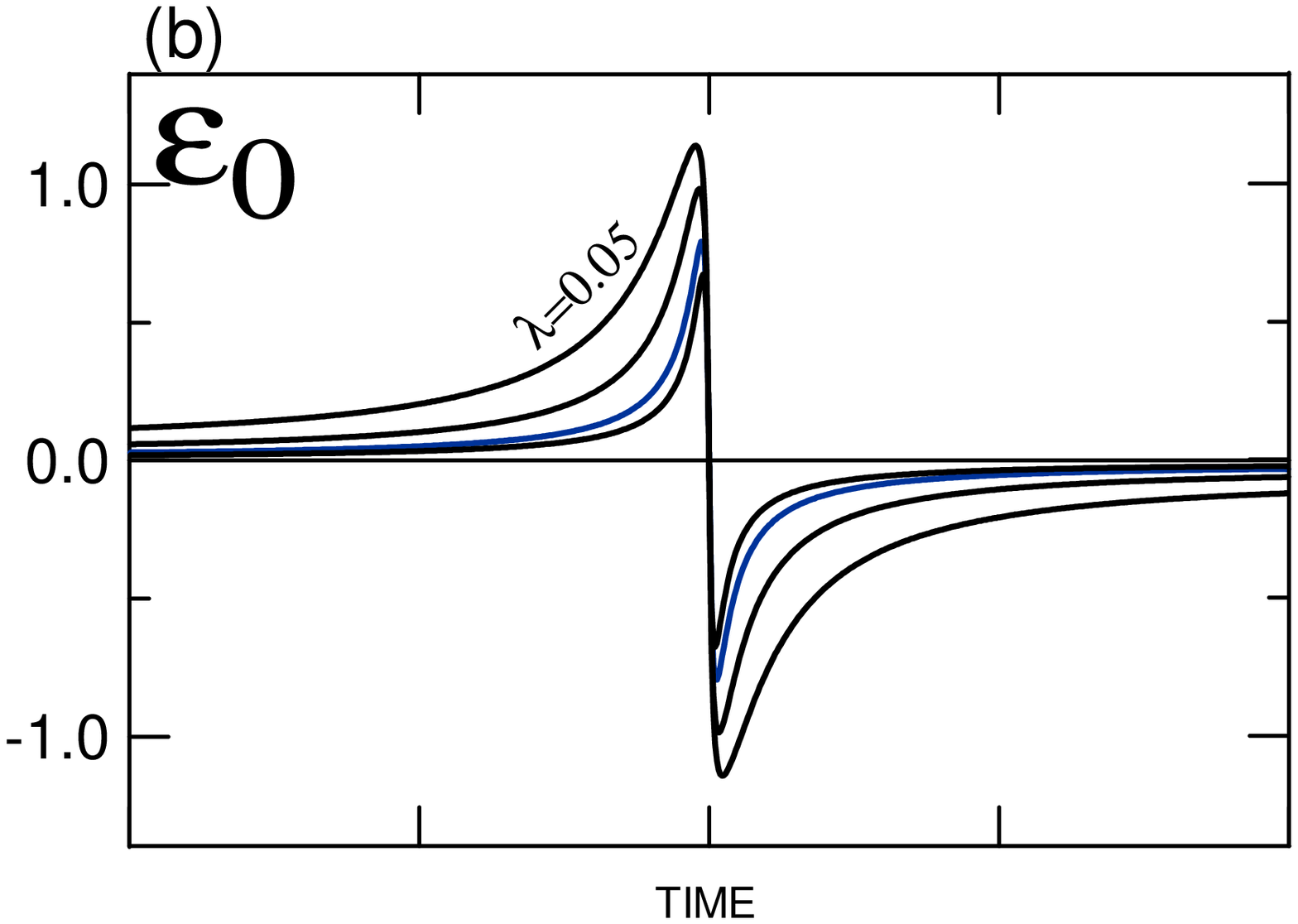}\hspace{2mm}
\includegraphics[height=4.9cm,clip=]{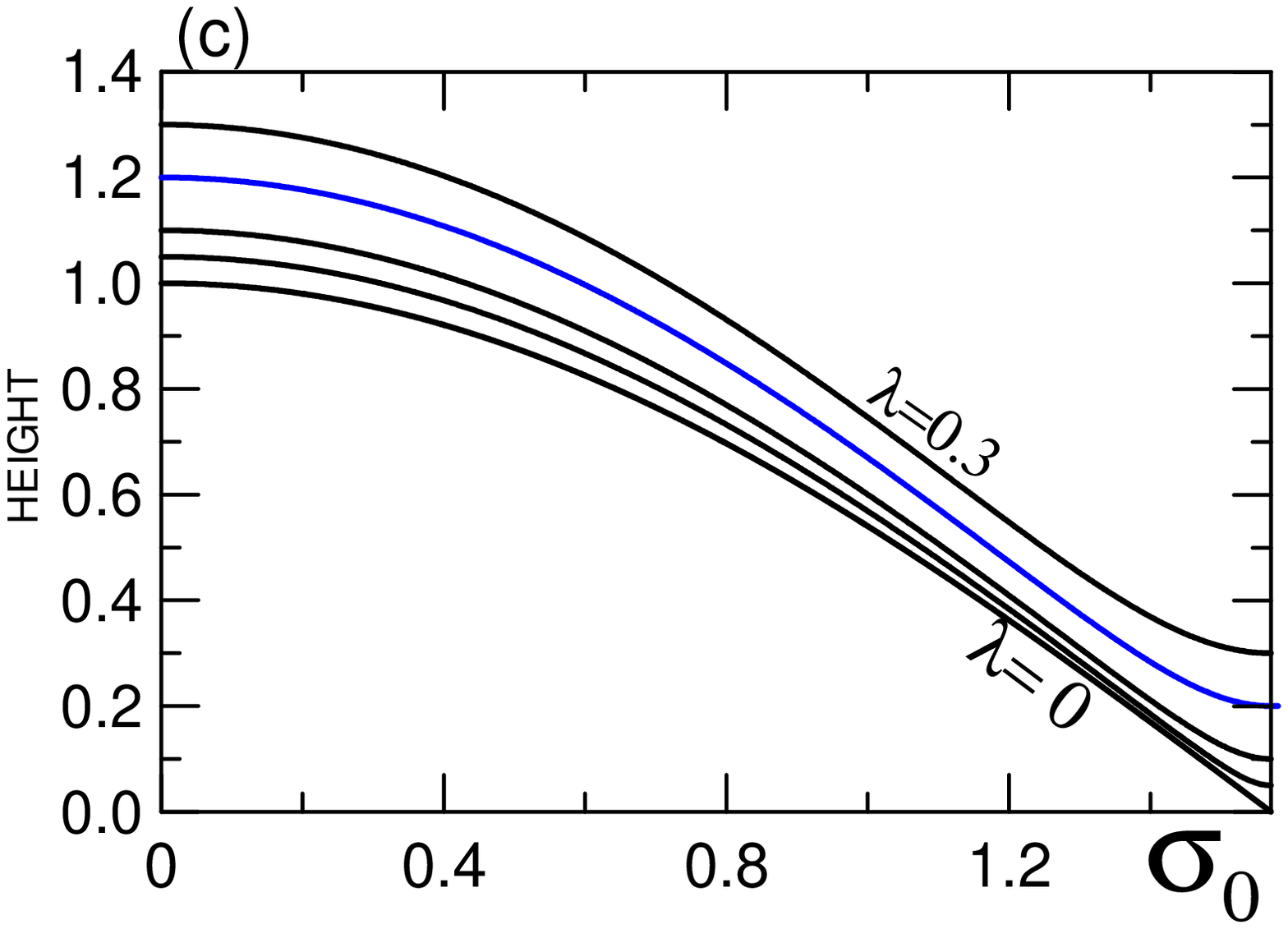}
}}
\caption{(a) Geodetic lag of the semi-diurnal tide as a function of $\sigma_0$. 
(b) Same as (a), but as a function of $\chi = \nu/\gamma$ $(\chi=\tan \sigma_0)$. 
(c) Time evolution of the geodetic tide lag when the frequency of the semi-diurnal tide crosses 0 and the tidal bulge changes of side with respect to the sub-$\tens{M}$ point. (d) Maximum height of the tide in units $\frac{1}{2}R\epsilon_\rho$. The blue lines correspond to $\lambda=0.2$ (Earth).}
\label{fig:h}
\end{figure}

\subsection{The geodetic lag}

The maximum tide height (i.e., the maximum of $\delta\zeta$) is, now, no longer reached at $\alpha=\sigma_0/2$ as the creep tide, but at
\beq
\alpha=\frac{1}{2} \ep_0
\endeq
where
\beq
\ep_0 =\arctan \frac{\sin 2\sigma_0}{1+2\lambda+\cos 2\sigma_0}.
\endeq

This function is shown in fig. \ref{fig:h}(a). We see that, as far as $\lambda \ne  0$, $\ep_0 \rightarrow 0$ when $\sigma_0 \rightarrow \frac{\pi}{2}$, that is, when $\gamma\rightarrow 0$. As a bonus, we also have near the rigid limit (i.e. near $\sigma_0=\frac{\pi}{2}$) $\ep_0$ decreasing when $\sigma_0$ increases, that is, when the frequency $\nu$ increases. This is exactly the behavior that is being advocated by Efroimsky and collaborators (Efroimsky and Lainey, 2007; Efroimsky and Williams, 2009; Castillo-Rogez et al. 2011; Efroimsky, 2012) for the Earth and the planetary satellites. 
This is best seen if we plot these curves using as independent variable the semi-diurnal frequency $\chi$ (in units of $\gamma$). See fig. \ref{fig:h}(b).  
We see that for $\chi$ small, $\varepsilon_0$ grows with $\chi$ (as in Darwin's theory). However, in this model,  it only grows up to reach a maximum and then decreases. We may compare these curves to the curve $Q(\chi)$ presented in fig. \ref{fig:Q} and remind of the popular use of the relation $Q=1/\varepsilon_0$ in Darwinian theories. However, the matching is very imperfect: The minimum of $Q$ happens for $\chi=1$ while the maximum of $\varepsilon_0$ happens for values of $\chi$ larger than 1, which increases indefinitely as $\lambda$ tends to zero.

It is worth recalling that one of the difficulties created by the assumption that the actual tide lag is proportional to a negative power of the frequency happens  when the frequency changes of sign. If no additional assumption is done, we just have a singularity with the tide lag tending to infinity or, at least, abrupt jumps between positive and negative values. This is not the case with the solution that results from the superposition of the elastic and creep tides. In this case, the transition from one side to another is smooth. The tide angle increases when the frequency decreases up to reach a maximum; after that point it quickly decreases up to cross zero with a finite derivative. The behavior in the negative side is just symmetrical (see fig\ref{fig:h}(c) ).

In addition, the agreement of fig \ref{fig:h}(c) with the red curve of Figure 1 of Efroimsky (2012) is noteworthy. Efroimsky's curve corresponds to the left half of fig\ref{fig:h}(c). The only differences are the reversed direction (fig. \ref{fig:h}(c) is drawn assuming that the frequency is decreasing as the time increases) and the use, by Efroimsky, of logarithmic scales. The two figures show exactly the same behavior notwithstanding the fact that they arise from two completely different models.

\subsection{Tide maximum height}\label{sec:height}
The maximum height of the creep tide (after the transient phase is over) can be determined as a function of the semi-diurnal frequency and the relaxation factor. It is given by the value of $h=\zeta-R^\prime$ at $ \alpha= \alpha_{max}$, that is 
\beq
h_{max}=\zeta(\alpha_{max})-R^\prime.
\endeq

Hence, for $\gamma t\gg 1$ and $\widehat\theta^*=90^\circ$ (i.e. at the equator of $\tens{m}$), in this approximation:
\beq
h_{max} 
=\frac{1}{2}R \epsilon_\rho \cos\sigma_0
=\frac{\frac{1}{2}R \epsilon_\rho}{\sqrt{1+\tan^2 \sigma_0}}
=\frac{\frac{1}{2}R\gamma \epsilon_\rho}{\sqrt{\gamma^2+\nu^2}}.
\endeq
This result is the same obtained by Darwin (1879) for the tides of a viscous spheroid, when inertia is neglected. 

In the case of one solid body, $\gamma=0$ and the creep tide height vanishes.
In the case of an inviscid body, $\gamma\rightarrow\infty$,  $\sigma_0 \rightarrow 0$, and we get the value 
$h_{max} =\frac{1}{2}R\epsilon_\rho$.

It is worth noting that the latest results appear clearly in the solution of the differential equation (eq. \ref{eq:nonelastic}) when it is written as 
\beq
\zeta={\cal{C}}e^{-\gamma t} + R^\prime + \frac{1}{2} R\epsilon_\rho^\prime \cos\sigma_0
 \cos (2\alpha - \sigma_0).
\endeq

When an instantaneous elastic tide is added to the creep tide, the maximum height of the geodetic (or composite) tide is the value of the function $\delta\zeta_{\rm eq}$ (eqn. \ref{eq:geodetic}) at its maximum:
\beq
h_{\rm max} = \frac{1}{2}R\epsilon_\rho \sqrt{\lambda^2+(1+2\lambda)\cos^2 \sigma_0}.
\endeq
Hence, it is almost unchanged when the viscosity is low ($\nu \ll \gamma$ i.e. $ \sigma_0 \ll 1$), but the difference becomes significant in the case of viscous bodies, when $\sigma_0$ is high. The relative value of the maximum height of the actual tide is shown in fig. \ref{fig:h}(d). In that figure, the unit is the maximum height of the equilibrium spheroid ($\frac{1}{2}R\epsilon_\rho $). One may note that when $\gamma\ll \nu$, $\sigma_0 \rightarrow \pi/2$, the height of the creep tide tends to zero and the maximum height of the composite tide is the maximum height of the instantaneous elastic tide: $\frac{1}{2}\lambda R\epsilon_\rho $. 
 
It is important to emphasize that the frequency-dependent height of the tide has not been taken into account in the majority of modern tide theories (Jeffreys, 1961; MacDonald, 1964; Kaula. 1964; Singer, 1968; Mignard, 1979; Hut, 1981; Murray and Dermott, 1999; Laskar and Correia, 2004).
In these theories, Love's elasticity theory is used to calculate the potential of the tidally deformed body, which is proportional to the frequency-independent Love number $k_2$ (or, in higher-orders, made of spherical harmonics proportional to the frequency-independent $k_j$, $j=2,...$). They consider the tides on an elastic body and, to take into account the imperfect elasticity, just delay the potential introducing by hand a phase lag. 

\subsection{Dynamical consequences of the elastic tide}

This empirical introduction of an elastic tide is necessary to have the theory conform to the measurements of Earth's bodily tides. It is however important to stress the fact that, being elastic, it does not present a lag.
The major axis of the prolate spheroid corresponding to the elastic tide is permanently oriented towards $\tens{M}$.

It then follows,
\beq
F_{1{\rm el}} \speq  
-\frac{4k_fGMmR^2\lambda\epsilon_\rho^\prime}{5r^{4}}, \hspace*{2cm} F_{2{\rm el}} \speq
F_{3{\rm el}} \speq 0.
\endeq

The force due to the empirical elastic tide is radial and its torque is equal to zero.
Therefore it does not contribute to the variation in the rotation discussed in section \ref{sec:rot}.
Its contribution to the dissipation also vanishes since the term added to $\mathbf{F}\mathbf{v}$ is proportional to $r^{-7}\sin v$ whose time average is zero. So, its introduction does not affect any of the results presented in previous sections

\section{Conclusions}

We will be brief in these conclusions.
The rheophysical theory presented in this paper is yet incipient to justify a long discussion. 
However, even at this early stage, some results are noteworthy.

The first result concerns the problem that served as motivation for this investigation: the so-called pseudo-synchronous stationary rotation.
In this theory, it is given by a law that depends on the viscosity of the body. 
In the limit $\gamma \rightarrow \infty$ (or $\eta \rightarrow 0$), the body behaves like a perfect fluid and the pseudo-synchronous stationary rotation happens exactly as in classical Darwin's theory.  However, when $\gamma \ll  n$ (e.g. in rocky bodies), no matter if the eccentricity is large or small, the stationary rotation is close to the true synchronous rotation. In the case of rocky bodies, the monthly/annual tide (due to eccentricity) does not create the strong torque responsible for the super-synchronicity of fluid bodies. 

In the case of natural satellites, the reophysical theory allows us to obtain the excess of the rotation period and to determine the ``length-of-day" (a.k.a synodic rotation period). The results are in agreement with the observed values. They were obtained without adding any torque due to some ``permanent" equatorial asymmetry, which may indeed exist, but is not necessary to counterbalance the tidal torque.

The second result concerns dissipation. Energy dissipation appears proportional to 
$\displaystyle(\frac\nu\gamma+\frac\gamma\nu)^{-1}$. 
The maximum dissipation of a tide component is reached when its frequency equals $\gamma$ and decreases symmetrically when $\nu$ is different of $\gamma$ no matter if larger or smaller. When $\nu \ll \gamma $ the energy dissipation is proportional to $\nu$. This is what happens in giant planets and stars. When $\nu \gg \gamma $, the energy dissipation is inversely proportional to $\nu$. This is the behavior of dissipation in rocky bodies as extensively discussed by Efroimsky and Lainey (2007), Efroimsky and Williams (2009), Castillo-Rogez et al. (2011) and Efroimsky (2012). 
We have taken some care writing these sentences to avoid privileging $\gamma$ or $\nu$. In fact, the relaxation factor $\gamma$ plays a role of critical frequency and one body may behave in different ways under the action of two tidal components of different frequencies if one is larger and the other is smaller than the critical frequency $\gamma$. In table 1 we have included in one column the inverse of $\gamma$, to give a clearer idea of the location of this bifurcation, which, in  some cases, is not very far from the periods of the actual stresses acting on the body.

At this point, I would like to stress that we have not used the quality factor $Q$ in the present theory. In standard theories, the quality factor $Q$ is an ambiguously defined parameter. It is defined using the semi-diurnal tide in the case of a freely rotating body, but using the monthly/annual  tide in the case of a pseudo-synchronous companion. In classical applications, these two cases are very distinct and we can handle the two different definitions. However, in the case of high-eccentricity exoplanets, this separation no longer occurs. The dissipation is shared in comparable parts by the semi-diurnal and the monthly/annual tide and we get different values of $Q$ following we consider one tide component or another. 
The low-eccentricity equivalence formulas relating $Q$ to $\gamma$ and the values listed in table 1 only appear in this paper because it is important to have a bridge between standard theories and this new one. However, it is necessary to stress the fact that these formulas are just numerical bridges valid for low eccentricities and for the rotational states assumed to establish them. Strictly speaking, a universal relation between $Q$ and $\gamma$ does not exist. We also note that the tidal Love number $k_2$ has not been used in the equations of the creeping tide.

Section \ref{examples} presents a short inventory of bodies in the Solar System and extrasolar.
For each of them, the values of the relaxation factor $\gamma$ and the uniform equivalent viscosity are derived on the grounds of the results obtained for them with standard theories. We stress the fact that the results shown allow the studied bodies to be divided into two groups. One group, including the planetary satellites and the terrestrial planets, in which the dissipation decreases when the frequency increases (Efroimsky-Lainey regime), and the other, including giant planets, hot Jupiters and stars, following Darwin's regime in which dissipation increases with the frequency.

At last, the Appendix discusses some accuracy problems involved in tidal evolution theories.
The warning included in  FRH stating that long series expansions by themselves are not improvements of a physical theory has been several times misunderstood and criticized (e.g. Leconte et al. 2011). However, we repeat it.
Very-high-order expansions did create, in the past, the feeling that the results of standard Darwin theory were exact notwithstanding the fact that we cannot yet say that we understand the Physics ruling bodily tides.
Indeed, we guess that the results of the standard Darwin theory for the pseudo-synchronous stationary rotation of close-in companions, independent of the viscosity of the body, cannot be correct, even when written using endless series.

On the negative side, the main question is that the rheophysical theory failed to give the actual shape of the bodily tide observed in the Earth. This failure forced us to admit the existence of a superposed elastic tide, which affects the shape of the observed tide but which is torque free and thus does not change the results of the rheophysical theory in what concerns the rotation of the bodies and the average energy dissipation. This superposition of the creeping tide and an elastic tide is a question that certainly needs to be stated in terms of Physics. 

In addition, we can devise many other points in which the theory needs to be improved. We list a few of them: We have considered only homogeneous bodies; we have considered a non-rotating model for the equilibrium tide (a Jeans spheroid); we have considered only the planar (two-dimensional) problem; we have disregarded inertia and the non-linearity of actual creep laws; we have assumed tidal deformations small enough so as to allow a superposition principle to be used in the calculation of the potential spanned by the tidally deformed body, etc. In that sense, the presented theory is yet a proposed model. 
Each of the cited points deserves now to be taken into consideration, and shall be taken into consideration if we want to apply the theory to bodies as intricate as some planetary satellites. One positive point to be mentioned is that the theory opens the way for the construction of very complex models and to adopt laws more complete than the Newtonian creep law.
Once the basic equations are given, they can be solved numerically.
This means that we may adopt physical models as complex as necessary, no matter if their equations can be solved analytically or not.

At last, we have to emphasize that the approach developed in this paper is a new and complete theory of the bodily tide problem, whose results derive from only one physical law: the Newtonian creep. There are no \textit {ad hoc} lags plugged by hand as in the standard theories. 
The constants $\sigma_i$ appearing in the arguments of the trigonometric functions, which can be physically interpreted as delays, are well determined parameters of the exact solution of simple ordinary differential equations. 
The results show some discrepancies with respect to the classical theories, but also some coincidences. The way in which creep and elastic tides combine to give rise to geodetic lags increasing when the frequency decreases, for hard bodies is appealing and may serve as a justification to the modern theories of Efroimsky and collaborators. In what concerns the discrepancies, it is important to know if they refer to observable phenomena or not. When observational data do not exist, it is impossible to know whether a proposed theory is or not correct. 

\begin{acknowledgements}
I wish to acknowledge the visiting fellowship granted by the Isaac Newton Institute for Mathematical Sciences, University of Cambridge (UK), where the basic ideas of this theory were first developed. I thank
Michael Efroimsky for our fruitful endless discussions about tides, rheology and dissipation, 
Rudi Dvorak and the Wien Universit\"at, where parts of this paper were written,
and Dr. G.Torres for kindly sending me the data used to construct fig. \ref{fig:binary}.
I am also indebted to Adri\'an Rodr\'{\i}guez and Hugo Folonier for having pointed out many mistakes in the previous versions of this paper.
This investigation was supported by CNPq, grants 302783/2007-5 and 306146/2010-0.

\end{acknowledgements}

\appendix

\section{Higher-order approximations}\label{higherorders}

The eccentricity functions introduced in section \ref{fullmodel} are the Cayley expansions for the solutions of the Keplerian motion (see Cayley, 1861; da Silva Fernandes, 1996) corresponding to the entries $(r/a)^{-3} [\sin|\cos] 2f$ in Cayley's tables. 

The approximations adopted in Sec. \ref{fullmodel} may be enough for most purposes. 
They are certainly enough for low eccentricities. For high eccentricities, however, 
a different approach is necessary. 
Cayley expansions include in their derivation the Taylor expansion of the solution of Kepler's equation, whose convergence radius is $e^*=0.6627434...$ (see Wintner, 1941). 
Some simple comparisons allow us to see that for eccentricities of the order of 0.5, 
or even less, the results are not accurate. 

\begin{table}[h]
\caption{Comparison of the calculated values of $E_{2,k}(e)$ for $e=0.4$}
\begin{tabular}{rrr}
 $k$ & \phantom{x}eq.(\ref{Fourier}) & series to $e^7$ \\
\hline\\
 -4 & 0.54807 &  0.49883\phantom{e-}\\
 -3 & 0.74504 &  0.74853\phantom{e-}\\
 -2 & 0.91810 &  0.92062\phantom{e-}\\
 -1 & 0.94573 &  0.94571\phantom{e-}\\
 0  & 0.62029 &  0.62030\phantom{e-}\\
 1 & -0.19615 & -0.19615\phantom{e-}\\
 2 & 0.\phantom{00000}&  0.\phantom{00000}\phantom{e-}\\
 3 & 0.00150 & 0.00150\phantom{e-}\\
 4 & 0.00120 & 0.00119\phantom{e-}\\
\end{tabular}
\label{tab:compara}
\end{table}

\begin{figure}[t]
\centerline{\hbox{
\includegraphics[height=4.5cm,clip=]{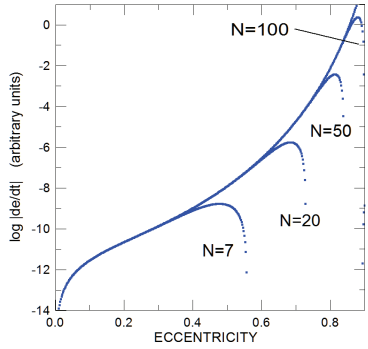}}}
\caption{$\log |de/dt|$ obtained with the series of eqn. (\ref{eq:doteav}) truncated at $N= $ 7, 20, 50 and 100. Arbitrary units}
\label{fig:dedt}       
\end{figure}

Since the equations appearing in this paper are written in terms of one of the families of Cayley expansions, the power series may be substituted by eq. \ref{Fourier-ell} in the form 
\beq\label{Fourier}
E_{2,k}(e)=\frac{1}{2\pi\sqrt{1-e^2}}\int_0^{2\pi}\frac{a}{r}\cos\big(2v+(k-2)\ell\big)\ dv.
\endeq
We may note that the function under the integral involves both the mean ($\ell$) and the true ($v$) anomalies, but can be easily handled numerically. The only difference with eq. (\ref{Fourier-ell}) is the preference in having the integration done over the true anomaly instead of being done over the mean anomaly (via the classical expression $r^2 dv=a^2\sqrt{1-e^2}d\ell$) to avoid having to solve the Kepler equation, which is operationally expensive. 

The integral solves the problem of the accuracy in the determination of the Cayley coefficients, but not that of the Fourier series giving the variation of the elements. 
To obtain approximations valid in high eccentricities, it is necessary to use expansions not affected by the singularities of the Keplerian motion (see Ferraz-Mello and Sato, 1989).
However, one preliminary guess can be done comparing different approximations.
Fig. \ref{fig:dedt} shows the value of $|\dot{e}|$ for close-in companions in pseudo-synchronous stationary rotation, in arbitrary units, considering the sum of the eccentricity-dependent terms of Eqn. \ref{eq:doteav} truncated at $N=$ 7, 20, 50 and 100, respectively. 
The abrupt changes of the curves with respect to the next one show a limiting eccentricity in each case. For instance the series truncated at $N=7$ (the usual cutoff of Cayley series) give wrong results for eccentricities above 0.4. 
It is important to stress that the calculations showed in fig. \ref{fig:simu}, concerning the tidal evolution predicted by the standard Darwin theory are not affected by truncation effects since for that sake we have used Mignard's formulation whose basic equations are given in closed form (Mignard, 1979) without resorting to any expansions.

\end{document}